\documentclass[a4paper,numberwithinsect,USenglish]{lipics-v2021}

\title{Existential Second-Order Logic Over Graphs: Parameterized~Complexity}

\titlerunning{ESO-Logic Over Graphs: Parameterized Complexity}

\author{Max Bannach}{European Space Agency, Advanced Concepts Team, Noordwijk,
  The Netherlands}{max.bannach@esa.int}{https://orcid.org/0000-0002-6475-5512}{}
\author{Florian Chudigiewitsch}{Universität zu Lübeck, Germany}{fch@tcs.uni-luebeck.de}{https://orcid.org/
0000-0003-3237-1650}{}
\author{Till Tantau}{Universität zu Lübeck,
Germany}{tantau@tcs.uni-luebeck.de}{}{}

\authorrunning{M. Bannach and F. Chudigiewitsch and T. Tantau}

\Copyright{M. Bannach and F. Chudigiewitsch and T. Tantau} 

\ccsdesc[500]{Theory of computation~Finite Model Theory}
\ccsdesc[500]{Theory of computation~Complexity theory and logic}
\ccsdesc[500]{Theory of computation~Fixed parameter tractability}
\ccsdesc[500]{Theory of computation~W hierarchy}
\keywords{existential second-order logic, graph problems, parallel
  algorithms, fixed-parameter tractability, descriptive complexity}

\category{} 

\relatedversion{} 

\acknowledgements{We thank Marcel Wienöbst for fruitful discussions and helpful comments on an earlier draft.}

\nolinenumbers 

\hideLIPIcs  

\EventEditors{John Q. Open and Joan R. Access}
\EventNoEds{2}
\EventLongTitle{42nd Conference on Very Important Topics (CVIT 2016)}
\EventShortTitle{CVIT 2016}
\EventAcronym{CVIT}
\EventYear{2016}
\EventDate{December 24--27, 2016}
\EventLocation{Little Whinging, United Kingdom}
\EventLogo{}
\SeriesVolume{42}
\ArticleNo{23}

\theoremstyle{plain}
\newtheorem{problem}[theorem]{Problem}
\newtheorem{fact}[theorem]{Fact}

\usepackage{booktabs}

\newcommand\Para{\mathrm{para\text-}}

\newcommand\Class[1]{%
  \mathchoice%
  {\text{\normalfont\small$\mathrm{#1}$}}%
  {\text{\normalfont\small$\mathrm{#1}$}}%
  {\text{\normalfont$\mathrm{#1}$}}%
  {\text{\normalfont$\mathrm{#1}$}}%
}
\newcommand\PClass{\mathrm{p\text{\normalfont-}}\Class}

\newenvironment{parameterizedproblem}%
{%
  \leavevmode\nobreak\par
  \begin{list}%
    {}%
    {%
      \def\labelstyle{\itshape}
      \setlength{\topsep}{0pt}%
      \settowidth{\labelwidth}{\labelstyle Parameter:}%
      \setlength{\leftmargin}{\labelwidth}%
      \addtolength{\leftmargin}{\labelsep}%
      \setlength{\itemsep}{0pt}%
      \setlength{\parsep}{0pt}%
    }%
      \def\instance{\item[\labelstyle Instance:]}%
      \def\parameter{\item[\labelstyle Parameter:]}%
      \def\question{\item[\labelstyle Question:]}%
    }%
    {%
  \end{list}%
}

\newcommand{\Lang}[1]{\text{\normalfont\textsc{#1}}}
\newcommand{\PLang}[2][]{\mathrm{p}_{#1}\Lang{-#2}}

\newcommand\coloneq{\mathrel{\raise.4pt\hbox{:}{=}}}
\newcommand\eqcolon{\mathrel{{=}\raise.4pt\hbox{:}}}

\usepackage{tikz}
\usetikzlibrary{graphs, arrows.meta, positioning, backgrounds, shapes, decorations.pathmorphing,}
\tikzset{
  every picture/.style={semithick},%
  node/.style={draw,circle,minimum size=5mm,inner
    sep=0.5pt,font=\footnotesize,fill=white},%
  small node/.style={node,minimum size=3.5pt,inner sep=0pt,outer
    sep=0pt,font=\tiny},%
  on/.style={fill=white,inner sep=-.4pt,circle}
}

\newcommand\patterntwo[8]{%
  \smash{%
  \tikz[baseline=-.5ex,any/.style={black!25,on/.style={inner sep=-.4pt,circle}}]{
    \node [node, minimum size=3mm] at (1,0) (w) {};
    \node [node, minimum size=3mm,fill=black] at (0,0) (b) {};
    \draw (w) edge [out=25, in=-25,looseness=12,#1] node  [on,near start,anchor=center] {\scriptsize#2} (w);
    \draw (b) edge [out=155, in=205,looseness=12,#3] node  [on,near
    start, anchor=center] {\scriptsize#4} (b);
    \draw (b) edge [bend left=20,#5] node [on] {\scriptsize#6} (w);
    \draw (w) edge [bend left=20,#7] node [on] {\scriptsize#8} (b);
  }}%
}
\newcommand\patterntwomixed[4]{%
  \smash{%
  \tikz[baseline=-.5ex,any/.style={black!25,on/.style={inner sep=-.4pt,circle}}]{
    \node [node, minimum size=3mm] at (1,0) (w) {};
    \node [node, minimum size=3mm,fill=black] at (0,0) (b) {};
    \draw (b) edge [bend left=20,#1] node [on] {\scriptsize#2} (w);
    \draw (w) edge [bend left=20,#3] node [on] {\scriptsize#4} (b);
  }}%
}

\def\AtMost{\textcolor{black!50}{p \rlap{\,$\preceq$}}}
\def\AtLeast{\textcolor{black!50}{\llap{$\preceq$\,} p\rlap.}}
\def\Between{\textcolor{black!50}{\llap{$\preceq$\,} p\rlap{\,$\preceq$}}}

\usepackage{tcs-automove}

\begin{document}

\maketitle

\begin{abstract}
  By Fagin's Theorem, NP contains precisely those problems that
  can be described by formulas starting with an existential second-order
  quantifier, followed by only first-order quantifiers (\textsc{eso}
  formulas). Subsequent research refined this result, culminating in
  powerful theorems that characterize for each possible sequence of first-order quantifiers how
  difficult the described problem can be. We transfer this line of
  inquiry to the \emph{parameterized} setting, where
  the size of the set quantified by the second-order quantifier is the
  parameter. Many natural parameterized problems can be described in
  this way using simple sequences of first-order quantifiers:
  For the clique or vertex cover problems, two universal first-order
  quantifiers suffice (``for all pairs of vertices \dots\ must
  hold''); for the dominating set problem, a universal followed 
  by an existential quantifier suffice (``for all vertices, there is a
  vertex such that \dots''); and so on. We present
  a complete characterization that states for each possible
  sequence of first-order quantifiers how high the parameterized
  complexity of the described problems can be. The uncovered dividing
  line between quantifier sequences that lead to tractable versus
  intractable problems is distinct from that known from the classical
  setting, and it depends on whether the parameter is a lower bound on,
  an upper bound on, or equal to the size of the quantified set.
\end{abstract}

\section{Introduction}

The 3-coloring problem is to decide, given an
undirected simple graph, whether there exist
three sets $R$, $G$, and~$B$ (the red, green, and blue
vertices) such that any two vertices $x$ and $y$ connected by an edge have
different colors; or in logical terms:
\begin{align}
  \exists R \exists G \exists B\, \forall x \forall y \smash{\Bigl(}&(Rx \lor Gx
  \lor Bx) \land{} \notag\\[-2pt]
  &\bigl(x\sim y \to \neg\bigl((Rx \land Ry) \lor (Gx \land Gy) \lor
  (Bx \land By)\bigr)\bigr)\smash{\Bigr)}. \label{eq-3col}
\end{align}
This formula is an \emph{existential second-order} formula, meaning
that it starts with existential second-order quantifiers ($\exists
R\exists G \exists B$) followed by first-order quantifiers ($\forall x
\forall y$) followed by a quantifier-free part. We can succinctly
describe which quantifiers are used in such a prefix by using
``$E_1$'' for a (monadic, hence the ``$_1$'') existential second-order
quantifier and ``$e$'' and ``$a$'' for 
existential and universal first-order quantifiers, respectively. The
resulting \emph{quantifier pattern} of the above formula is then
$E_1E_1E_1aa$; and (monadic) existential second-order formulas are
 formulas with a prefix in~$E_1^*(ae)^*$. 
 It is no coincidence that an $\Class{NP}$-complete problem can be
 described using the quantifier pattern $E_1E_1E_1aa$: Fagin's
 Theorem~\cite{Fagin74} states that a problem lies in $\Class{NP}$ iff
 it can be described by a formula with a pattern in $E_i^*(ae)^*$ for
 some arity~$i$. However, the example shows that the pattern
 $E_1E_1E_1aa$ already suffices to describe an $\Class{NP}$-complete
 problem and a closer look reveals that so does $E_1E_1aa$. In
 contrast, formulas with the pattern~$E_iaa$ can only describe
 problems decidable in~$\Class{NL}$ (regardless of the arity~$i$ of
 the quantified relation variables). Such observations have sparked an
 interest in different quantification patterns' power. The
 question was answered by Gottlob, Kolaitis, and
 Schwentick~\cite{GottlobKS04} in the form of a dichotomy (``can only
 describe problems in $\Class P$'' versus ``can describe an
 $\Class{NP}$-complete problem'') and later in a refined form by
 Tantau~\cite{Tantau15}, where the described problems in~$\Class P$
 are further classified into ``in~$\Class{AC}^0$'' or
 ``$\Class L$-complete'' or ``$\Class{NL}$-complete''.

While in the formula for 3-colorability it was only necessary that
three sets of colors \emph{exist,} for many problems the \emph{size}
of these sets is important. Consider: 
\begin{align}
  \phi_{\text{clique}} &= \exists^{\ge} C \, \forall x \forall y \bigl((Cx
  \land Cy) \to x \sim y\bigr), \label{eq:clique} \\
  \phi_{\text{vertex-cover}} &= \exists^{\le} C \,
  \forall x \forall y \bigl(x \sim y \to (Cx \lor Cy)\bigr), \label{eq:vc} \\
  \phi_{\text{dominating-set}} &= \exists^{\le} D \, \forall x \exists
  y \bigl(Dy \land (x = y \lor x \sim y)\bigr). \label{eq:domset}
\end{align}
where the second-order quantifiers $\exists^{\ge}$ and $\exists^{\le}$ ask
whether there exists a set of size at least or at most some parameter value~$k$
such that the rest of the formula holds. These formulas show that we can
describe the clique problem using a formula with the succinctly written pattern
$E^{\ge}_1 aa$ (and also $E_1^=aa$); the vertex cover problem using $E^{\le}_1
aa$ (and again also $E_1^= aa$); and the dominating set using $E^{\ge}_1 ae$
(and yet again also $E_1^=ae$). Readers will notice that the problems are some
of the most fundamental problems studied in that theory and lie in different
levels of the $\Class W$-hierarchy. The main message of the present paper is
that it is once more \emph{no coincidence that the quantifier patterns needed to
describe these problems differ} ($E^{\ge}_1 aa$ versus $E^{\le}_1 aa$ versus
$E^{\ge}_1 ae$): As done in \cite{GottlobKS04,Tantau15}, we will give a complete
characterization of the complexities of the problems that can be described using
a specific quantifier pattern. The well-known results that the (parameterized)
clique and dominating set problems are $\Class W[1]$-hard while the
(parameterized) vertex cover problem lies in $\Class{FPT}$ (in fact, in
$\Para\Class{AC}^0$) can now all be derived just from the syntactic structure of
the formulas used to describe these problems.

\begin{table}[htbp]
  \caption{Complete complexity classification of the weighted
    \textsc{eso} logic for a single weighted monadic second-order 
    quantification followed by first-order quantifiers with some
    pattern $p \in \{a,e\}^*$ (where $p \preceq q$ means that $p$ is a 
    subsequence of~$q$). The upper part (arbitrary structures) and
    lower part (basic graphs) are identical except for the
    patterns $E_1^\ge ae$ and $E_1^\le aa$, where they differ. Note
    that $\Para\Class{AC}^0 \subsetneq 
    \Para\Class{AC}^{0\uparrow} \subseteq \Para\Class P = \Class{FPT}$
    and $\Class{FPT} \cap \Class W[1]\text{-hard} =\emptyset$ is a
    standard assumption.}
  \label{table:summary-weighted-fd-fo}
  \begin{tabular}{llrcl}
    \toprule
    $\PClass{FD}(E_1^{=}p)$
    & $\subseteq \Para\Class{AC^{0}}$, when 
    &
    & $\AtMost$
    & $e^*a$.
    \\
    &  $\cap\ \Class{W}[1]\text{-hard} \neq \emptyset$, when
    & $ae$ or $aa$
    & $\AtLeast$
    &
    \\
    \midrule
    $\PClass{FD}(E_1^{\ge}p)$
    & $\subseteq \Para\Class{AC^{0}}$, when
    &
    & $\AtMost$
    & $e^*a$.
    \\
    & $\cap\ \Class{W}[1]\text{-hard} \neq \emptyset$, when
    & $ae$ or $aa$
    & $\AtLeast$
    &
    \\
    \midrule
    $\PClass{FD}(E_1^{\le}p)$
    & $\subseteq \Para\Class{AC^{0}}$, when
    &
    & $\AtMost$
    & $e^*a$.
    \\
    & $\not\subseteq \Para\Class{AC^{0}}$ but $\subseteq \Para\Class{AC^{0\uparrow}}$, when
    & $aa$
    & $\Between$
    & $e^*a^*$.
    \\
    &  $\cap \ \Class{W}[1]\text{-hard} \neq \emptyset$, when
    & $ae$
    & $\AtLeast$
    &
    \\
    \bottomrule
    \\[-1em]
    \toprule
    $\PClass{FD}_{\mathrm{basic}}(E_1^{=}p)$
    & $\subseteq \Para\Class{AC^{0}}$, when
    &
    & $\AtMost$
    & $e^*a$.
    \\
    & $\cap\ \Class{W}[1]\text{-hard} \neq \emptyset$, when
    & $ae$ or $aa$
    & $\AtLeast$
    &
    \\
    \midrule
    $\PClass{FD}_{\mathrm{basic}}(E_1^{\ge}p)$
    & $\subseteq \Para\Class{AC^{0}}$, when
    &
    & $\AtMost$
    & $e^*a$ or $ae$.
    \\
    & $\cap\ \Class{W}[1]\text{-hard} \neq \emptyset$, when
    & $aee$, $eae$, or $aa$
    & $\AtLeast$
    &
    \\
    \midrule
    $\PClass{FD}_{\mathrm{basic}}(E_1^{\le}p)$
    & $\subseteq \Para\Class{AC^{0}}$, when
    &
    & $\AtMost$
    & $e^*a$ or $aa$.
    \\
    & $\not\subseteq \Para\Class{AC^{0}}$ but $\subseteq \Para\Class{AC^{0\uparrow}}$, when
    & $aaa$
    & $\Between$
    & $e^*a^*$.
    \\
    & $\cap\ \Class{W}[1]\text{-hard} \neq \emptyset$, when
    & $ae$
    & $ \AtLeast$
    &
    \\
    \bottomrule
  \end{tabular}
\end{table}

\subparagraph*{Our Contributions.}

In this paper, we classify the complexity of the following classes
(formal definitions are given in
Section~\ref{section:describing-parameterized-problems}): Given a 
pattern $p \in \{a,e\}^*$ of first-order quantifiers, the classes
$\PClass{FD}(E_1^{=}p)$, $\PClass{FD}(E_1^{\le}p)$, and $\PClass{FD}(E_1^{\ge}p)$
contain all parameterized problems that can be described by formulas
with quantifier pattern $E^{=}_1 p$, or $E^{\ge}_1 p$, or~$E^{\le}_1
p$, respectively. The restriction to study just a \emph{single,
monadic, parameterized} \textsc{eso} quantifier is motivated by our earlier
observation that important and interesting problems of parameterized
complexity can be described in this way.
Our classification is complete in the sense that for every~$p$ we
either show that all problems in the class are fixed-parameter 
tractable (in these cases, we derive more fine-grained results by placing the
problems in $\Para\Class{AC}^0$ or $\Para\Class{AC}^{0\uparrow}$) or
there is a $\Class W[1]$-hard problem that can be described using the 
pattern.
Table~\ref{table:summary-weighted-fd-fo} lists the obtained bounds. In the
table, the classes with the subscript ``basic'' refer to the restriction to
undirected graphs without self-loops. As can be seen, for these graphs we get
slightly different complexity results. This is in keeping with the classical,
non-parameterized setting studied by Gottlob et al.~\cite{GottlobKS04}, where
results for basic graphs are often \emph{considerably} harder to obtain.
However, the complexity landscape we uncover in the present paper is different
from the one presented in~\cite{GottlobKS04} and \cite{Tantau15}: Although
certain patterns (like $p = ae$) feature prominently in the parameterized and
non-parameterized analysis, the dividing lines are different. To establish these
lines, we combine ideas used in the classical setting with different methods
from parameterized complexity theory, tailored to the specific problems we
study. The notoriously difficult cases from the classical setting (Gottlob et
al.~\cite{GottlobKS04} spend 34 pages to address the case $E_1^* ae$,
Tantau~\cite{Tantau15} spends several pages on $E_1 aa$) are also technically
highly challenging in the parameterized setting\iftcsautomove\ (and we give the
proofs only in the appendix)\fi.

Our research sheds new light on what difference it makes whether we want
solutions to have size \emph{exactly~$k$} or \emph{at most~$k$} or \emph{at
least~$k$.} To begin, equations~\eqref{eq:clique} and~\eqref{eq:domset} already
show that for individual problems (like clique) the maximization problem can be
hard while minimization is trivial (a single vertex is always a clique) and for
some problems (like dominating set) the opposite is true (the whole vertex set
itself is always a dominating set). Furthermore, from the perspective of
descriptive complexity, there is a qualitative difference between $\exists^=C$
and~$\exists^\le C$ on the one hand and $\exists^\ge C$ on the other: For
any~$k$, the first two can easily be expressed in normal \textsc{eso} logic
using $k$ first-order quantifiers binding the elements of~$C$, while
$\exists^\ge$ translates to $\exists C \exists x_1\cdots\exists x_k$ where the
$x_i$ bind the elements not in~$C$. Thus, $\exists^=$ and $\exists^\le$ only
allow us to express problems that are ``slicewise first-order'' and hence in
$\Class{XAC}^0 \subseteq \Class{XP}$, while already the slice for $k=0$ of
$\exists^\ge$ formulas can express $\Class{NP}$-complete problems for many
patterns. \emph{However,} we also prove a result for basic graphs for $p = ae$
that runs counter this ``tendency'' of $\exists^\ge$ to be harder
than~$\exists^\le$: While $\PClass{FD}_{\mathrm{basic}}(E_1^{\le} ae)$ contains
the $\Class{W}[2]$-hard dominating set problem,
$\PClass{FD}_{\mathrm{basic}}(E_1^{\ge} ae) \subseteq \Para\Class{AC}^0$.  


\subparagraph*{Related Work.}

Using logic to describe languages dates back all the way to Büchi's pioneering
work~\cite{Buechi1960} on the expressive power of monadic second-order logic
(which, over strings, describes exactly the regular languages). Switching from
monadic second-order logic to existential second-order logic yields Fagin's
Theorem~\cite{Fagin74}. Since then, the expressive power of fragments of this
logic was the subject of intensive research: Eiter et al.~\cite{EiterGG00}
studied the expressiveness of \Lang{eso}-patterns over strings; Gottlob et al.\
did so over graphs~\cite{GottlobKS04}; Tantau~\cite{Tantau15} refined the latter
results for subclasses of~$\Class{P}$. Taken together, these results give us a
complete complexity-theoretic classification of the problems resulting from any
\textsc{eso} quantifier pattern over strings, basic graphs, directed graphs,
undirected graphs, and  arbitrary structures.

Using logical fragments to characterize complexity classes is also standard
practice in parameterized complexity theory~\cite{FlumG06}, especially the power
of \textsc{mso} logic plays a prominent role, see for
instance~\cite{CourcelleE12}. In particular, characterizations of the levels of
the $\Class W$-hierarchy in terms of the number of quantifier alternations are
known~\cite{DowneyF99,FlumG06}, but -- to the best of our knowledge -- a
complete and exact analysis of the parameterized complexity of problems in terms
of the quantifier patterns describing them is new.

\subparagraph*{Organization of this Paper.}

Following a review of basic concepts and terminology in
Section~\ref{section:background}, we present our results on the power
of quantifier patterns of the forms $E^{\le}_1 p$, $E^{\ge}_1 p$, and $E^{=}_1 p$ for $p \in
  \{a,e\}^*$ in the subsections of
Section~\ref{section:weighted-fd} (arbitrary structures) and
Section~\ref{section:weighted-fd-basic} (basic graphs).

\tcsautomoveaddto{main}{
  \clearpage
  \appendix
  \section{Technical Appendix}
  In the following, we provide the proofs omitted in the main text. In
  each case, the claim of the theorem or lemma is stated once more for
  the reader's convenience. 
}

\section{Background in Descriptive and Parameterized Complexity}
\label{section:background}
 
\subparagraph*{Terminology for Graphs and Logic.}
A \emph{directed graph} (``digraph'') is a
pair $G = (V,E)$ where~$V$ is a \emph{vertex set} and $E \subseteq V
\times V$ an \emph{edge set}. An \emph{undirected graph}
is a pair $G = (V,E)$ such that $E \subseteq \bigl\{\{u,v\} \mid u,v \in
  V\bigr\}$. A \emph{basic graph} is an undirected graph that has no
self-loops, that is, where all edges have size~$2$. In this paper,
graphs are always finite.

We use standard terminology from logic and finite model theory, see
for instance \cite{EbbinghausFT94}. Let us point out some perhaps
not-quite-standard notation choices: Our vocabularies~$\tau$ (also known
as \emph{signatures}) contain \emph{only relation symbols} and we
write $\Lang{struc}[\tau]$ to denote the set
of all finite $\tau$-structures. For a first-order or second-order
$\tau$-sentence~$\phi$ (a formula without free variables), let
$\Lang{models}(\phi)$ denote the subset of $\Lang{struc}[\tau]$ of all
$\tau$-structures that are models of~$\phi$. As an example, we can
represent digraphs using the vocabulary $\tau_{\mathrm{digraph}} =
\{\sim^2\}$, containing a single binary 
relation symbol, and the class of digraphs is exactly
$\Lang{struc}[\tau_{\mathrm{digraphs}}]$. The formula $\phi = \forall x
  \forall y (x \sim y \to x \neq y)$ expresses that there are no
loops in a graph, that is, $\Lang{models}(\phi) = \{ G \mid G$ is a
digraph that has no self-loops$\}$. While an undirected graph $G =
  (V,E)$ is not immediately a $\tau_{\mathrm{digraph}}$-structure, we
can trivially ``turn it'' into a structure~$\mathcal G$ by setting the
universe to be~$V$ and setting $\sim^{\mathcal G} = \{(x,y) \mid
  \{x,y\} \in E\}$ and this structure is a model of
  $\phi_{\mathrm{undirected}} = \forall x \forall y (x \sim y \to y
  \sim x)$. The structures representing basic graphs are then
models of $\phi_{\mathrm{basic}} = \forall
  x\forall y(x\sim y \to (x \neq y \land y \sim x))$. As another example, the
class of all bipartite graphs equals
$\Lang{models}(\phi_{\mathrm{bipartite}})$ where
$\phi_{\mathrm{bipartite}}$ is the second-order formula $\exists X
  \forall u\forall v\bigl(u \sim v \to (Xu \leftrightarrow \neg Xv)\bigr)$
and $Xu$ is our shorthand for the less concise~$X(u)$.

As already sketched in the introduction, we can associate a
\emph{quantifier prefix pattern} (a word over the infinite alphabet
$\{E_1,E_2,E_3,\dots\} \cup \{e,a\}$), or just a \emph{pattern}, to
formulas of $\Lang{eso}$ logic by first writing them in prenex normal form
(quantifiers first, in a block) and then replacing each (existential)
second-order quantifier by $E_i$, where $i$ is the arity of the
quantifier, each universal first-order quantifier by~$a$, and each
existential first-order quantifier by~$e$. For instance, the pattern
of $\phi_{\mathrm{bipartite}}$ is $E_1aa$.

\subparagraph*{Describing Problems and Classes.}
\label{section:describing-parameterized-problems}

In the context of  descriptive complexity 
a \emph{decision problem~$P$} is a subset of $\Lang{struc}[\tau]$ that
is closed under isomorphisms. We say that \emph{$\phi$ describes $P$}
if $\Lang{models}(\phi) = P$. Moving on to classes, for a set~$\Phi$
of $\tau$-formulas, let $\Class{FD}(\Phi) \coloneq \{\Lang{models}(\phi)
\mid \phi \in \Phi\}$ denote the class of problems ``Fagin-defined''
by~$\Phi$. For a quantifier prefix pattern~$p$ let $\Class{FD}(p) \coloneq \{\Lang{models}(\phi)
\mid \phi$ has pattern~$p\}$, so \eqref{eq-3col} shows that
$\Lang{3colorable} \in \Class{FD}(E_1E_1E_1aa)$, and for a set $S$ of patterns let $\Class{FD}(S) = \bigcup_{p \in S}
\Class{FD}(p)$. In slight abuse of notation, we usually write
down regular expressions to denote sets $S$ of quantifier patterns:
For instance, Fagin's Theorem~\cite{Fagin74} can now be written as
``$\Class{NP} = \Class{FD}\bigl(E_2^*(ae)^*\bigr)$.''
Trivially, more quantifiers potentially allow us to express more
problems. Formally, let $p \preceq q$ denote that $p$ is a subsequence
of~$q$ (so $p$~can be obtained from~$q$ by, possibly, deleting some
letters). Then $\Class{FD}(p) \subseteq \Class{FD}(q)$. Also in slight
abuse of notation, we also write things like ``$p \preceq e^* a$'' to
indicate that $p \preceq q$ holds for some $q$ of the form $e^* a$.

Our analysis will show that restricting attention to basic graphs yields
particularly interesting results. For this reason, it will be convenient to
consider the introduced complexity classes restricted to basic graphs by adding a subscript
``basic'': For $\tau_{\mathrm{digraph}}$-formulas $\phi$, let
$\Lang{models}_{\mathrm{basic}}(\phi) = \Lang{models}(\phi) \cap
\Lang{models}(\phi_{\mathrm{basic}})$ and define
$\Class{FD}_{\mathrm{basic}}(\Phi)$ and $\Class{FD}_{\mathrm{basic}}(p)$ in the
obvious ways -- and similarly for the classes with the subscript ``undirected.''


When we move from classical complexity theory to parameterized
complexity, we assign to every \emph{instance} a \emph{parameter} that
measures an aspect of interest of that instance and that is hopefully
small for practical instances. A \emph{parameterized problem} is a set
$Q\subseteq\Lang{struc}[\tau]\times\mathbb{N}$ such that for every~$k$
the \emph{slice} $\{x \mid (x,k) \in Q\}$ is closed under
isomorphisms. In a pair $(x,k)$ we call $x$ the \emph{input} and $k$ the
\emph{parameter.} The usual goal in the field is to prove that a
problem is \emph{fixed-parameter tractable} (in $\Class{FPT}$) by
deciding $(x,k)\in^?Q$ in time
$f(k)\cdot |x|^{O(1)}$ for some computable function~$f$.
In the context of problems described by $\Lang{eso}$ formulas, a
natural parameter to consider is the \emph{size of the relations that
we can assign to the existential second-order quantifiers} and this
size is commonly called the \emph{solution weight}. As mentioned
earlier, problems like the vertex cover problem can be described
naturally in this manner: Consider the formula $\phi(X) = \forall u
\forall v \bigl(u \sim v \to (Xu \lor Xv)\bigr)$, where $X$ is a free
monadic second-order variable. Then for a graph $G = (V,E)$, viewed as
a logical structure $\mathcal G$,  and a set $C \subseteq V$ we have
$\mathcal G \models \phi(C)$ iff $C$ is a vertex cover of~$G$. Thus, $(\mathcal
G,k) \in \PLang{vertex-cover} =
\{(\mathcal G,k) \mid \mathcal G$~has a vertex cover of size~$k\}$ iff
there exists a set $C \subseteq V$ of size~$k$ such that
$\mathcal G \models \phi(C)$.

Formally, the second-order quantifiers $\exists^{\le}$,
$\exists^{=}$, and~$\exists^{\ge}$ have the following semantics: For
a structure~$\mathcal S$ with a universe~$U$, a non-negative integer~$k$, an $i$-ary
second-order variable~$X$, and a formula $\phi(X)$, we say that
\emph{$\mathcal S$ is a model of $\exists^{\le}X\, \phi(X)$
for parameter~$k$} and write $(\mathcal S,k) \models \exists^{\le}X\, 
\phi(X)$, if there is a set $C \subseteq U^i$ with $|C| \le  
k$ such that $\mathcal S \models \phi(C)$. A formula starting with a
$\exists^{\le}$ quantifier then gives rise to a
parameterized problem: Let
$\PLang{models}\bigl(\exists^{\le}X\,\phi(X)\bigr) = \{(\mathcal S,k)
\mid (\mathcal S,k) \models 
\exists^{\le}X \, \phi(X)\}$ and let $\PLang{}\Class{FD}(\Phi) \coloneq
\{\PLang{models}(\phi) \mid \phi \in \Phi\}$. The at-least and equal
cases are, of course, defined analogously. As an example, we have $\PLang{vertex-cover} \in
\PLang{}\Class{FD}(E_1^{\le}aa)$ since $\PLang{vertex-cover} =
\PLang{models}\bigl(\exists^{\le} X\, \forall u
\forall v \bigl(u \sim v \to (Xu \lor Xv)\bigr)\bigr)$.

\subparagraph*{Standard and Parameterized Complexity Classes}

Concerning standard complexity classes, we use standard definitions,
see for instance~\cite{AroraB09,Papadimitriou94}. In the context of
descriptive complexity theory, it is often necessary to address coding
issues (meaning the question of how words are encoded as logical
structures and \emph{vice versa}) -- but fortunately this will not be
important for the present paper.
Concerning parameterized complexity classes like $\Class{FPT} =
\Para\Class P$ or $\Class W[1]$, we also use standard 
definitions, which can be adapted to the
descriptive setting in exactly the same way as for classical
complexity classes (see for instance \cite{BannachST15,BannachT18}
for details) and encoding details will once more be unimportant.
The classes $\Para\Class{AC}^0$ and $\Para\Class{AC}^{0\uparrow}$ are
likely less well-known: We have $Q \in \Para\Class{AC}^0$ 
if there is a family $(C_{n,k})_{n,k \in \mathbb N}$ of unbounded
fan-in circuits of constant depth and size $f(k)\cdot n^{O(1)}$
for some computable function~$f$, such that for every $(\mathcal S,k)
\in \Lang{struc}[\tau] \times \mathbb N$ we have $(\mathcal S,k) \in
Q$ iff the circuit $C_{\mathrm{length}(\mathcal S),k}$ outputs $1$ on
input of (a suitably encoded) $\mathcal S$, where $\mathrm{length}(\mathcal S)$
is the length of the encoding of $\mathcal S$. For the class
$\Para\Class{AC}^{0\uparrow}$, the circuits may have depth~$f(k)$. We have $\Para\Class{AC}^0 \subsetneq
\Para\Class{AC}^{0\uparrow} \subseteq \Para\Class{P} =
\Class{FPT}$~\cite{BannachST15}.
In our proofs, two properties of the classes will be important: First, all of
them are (quite trivially) closed under 
$\Para\Class{AC}^0$-reductions. Second, for $\tau = (I^1)$, the signature with a
single unary relation symbol, we have
$\PLang{threshold} = \{(\mathcal S,k) \mid \mathcal S = (U,I^{\mathcal
  S}), |I^{\mathcal S}| \ge k\} \in \Para\Class{AC}^0$, that is, we
can ``count up to the parameter in $\Para\Class{AC}^0$.''
For more details on these classes, including
discussions of uniformity, see \cite{BannachST15,BannachT18,ChenF16}.

%
%

\section{Classifying Parameterized ESO Classes: Arbitrary~Structures}
\label{section:weighted-fd}

\tcsautomoveaddto{main}{\subsection{Proofs for Section~\ref{section:weighted-fd}}}

We now begin tracing the tractability frontier for the classes from
the upper part of Table~\ref{table:summary-weighted-fd-fo}:
$\PClass{FD}(E_1^{=}p)$, $\PClass{FD}(E_1^{\ge}p)$, and
$\PClass{FD}(E_1^{\le}p)$. Recall that for these classes we are given a
formula~$\phi$ starting with one of the monadic second-order quantifiers
$\exists^=$, $\exists^\le$, or $\exists^\ge$, followed by first-order
quantifiers with the pattern~$p$; and the objective is to show
\emph{upper bounds} of the form ``for all~$\phi$ with pattern~$p$ all
$\PLang{models}(\phi)$ lie in a certain class'' and \emph{lower
bounds} of the form ``there is a $\phi$ with pattern~$p$ such that
$\PLang{models}(\phi)$ contains a problem that is hard for a certain
class''. We dedicate one subsection to each of $\exists^=$,
$\exists^\le$, and $\exists^\ge$, each starting with the main theorem
and covering more technical parts of the proofs later. 

In this section, we allow arbitrary (finite, relational) structures, meaning that
the signature~$\tau$ can contain arbitrary relation symbols (but neither
constant nor function symbols), and our
upper bounds will hold for all such structures. However, for our
\emph{lower bounds} it will suffice to consider only
\emph{undirected graphs.} That is, the lower bounds for a pattern~$p$
will be of the form ``there is a $\phi$ with pattern~$p$ such that 
$\PLang{models}_{\mathrm{undirected}}(\phi)$ contains a hard
problem''. Interestingly, we can typically (\emph{but not always,} by the
results of Section~\ref{section:weighted-fd-basic}) replace
\emph{undirected} graphs by \emph{basic} graphs (undirected graphs
without self-loops) here.

\subsection{Solution Weight Equals the Parameter for Arbitrary Structures}

We start with the classification of $\PClass{FD}(E_1^{=}p)$, the first two
lines of Table~\ref{table:summary-weighted-fd-fo}:

\begin{theorem}[{Complexity Dichotomy for $\PClass{FD}(E_1^{=}p)$}]\label{theorem:arb-equal}
  Let $p \in \{a,e\}^*$ be a pattern.
  \begin{enumerate}
  \item $\PClass{FD}(E_1^{=}p) \subseteq \Para\Class{AC}^{0}$, if $p
    \preceq e^*a$.
  \item $\PClass{FD}(E_1^{=}p)$ contains a $\Class{W[1]}$-hard
    problem, if $aa\preceq p$ or $ae\preceq p$.
  \end{enumerate}
  Both items also hold for $\PClass{FD}_{\mathrm{undirected}}(E_1^{=}p)$ and even
  $\PClass{FD}_{\mathrm{basic}}(E_1^{=}p)$. 
\end{theorem}
The cases in the above theorem are exhaustive (so for every $p$ we
either have $p \preceq e^*a$ or we have  $aa\preceq p$ or $ae\preceq
p$). The theorem follows directly from the following lemma:
\begin{lemma}[{Detailed Bounds for $\PClass{FD}(E_1^{=}p)$}]\label{lemma:arb-equal}
  \hfil
  \begin{enumerate}
  \item\label{item:eq-e*a}  $\PClass{FD}(E_1^{=}e^* a) \subseteq
    \Para\Class{AC}^{0}$.
  \item\label{item:eq-aa} $\PClass{FD}_{\mathrm{basic}}(E_1^{=}aa)$
    contains a $\Class{W[1]}$-hard problem.
  \item\label{item:eq-ae} $\PClass{FD}_{\mathrm{basic}}(E_1^{=}ae)$
    contains a $\Class{W[2]}$-hard problem.
  \end{enumerate}
\end{lemma}

\begin{proof}
  Item~\ref{item:eq-e*a} is shown in Corollary~\ref{corollary:e-star-a}, which we prove later
  in this section. For item~\ref{item:eq-aa}, we already saw in
  equation~\eqref{eq:clique} that we can describe 
  the $\Class{W}[1]$-hard clique problem $\PLang{clique}$ using a
  formula $\phi_{\text{clique}}$ with pattern $E_1^{\ge}aa$. It was also
  already mentioned that replacing $\exists^\ge$ by $\exists^=$ yields
  the same problem and, thus, $\PClass{FD}_{\mathrm{basic}}(E_1^{=}aa)$ contains a
  $\Class{W}[1]$-hard problem. Similarly, for item~\ref{item:eq-ae}, replacing
  $\exists^\le$ by $\exists^=$ in equation~\eqref{eq:domset} shows we
  can describe the $\Class{W}[2]$-hard dominating set problem using
  an $E^{=}_1ae$ formula.
\end{proof}
To establish the upper bound (item~\ref{item:eq-e*a} of the theorem), we make
use of a well-known connection between weighted satisfiability in predicate
logic (problems in $\PClass{FD}(E_1^{=}e^*a)$ in our case) and weighted
satisfiability in propositional logic (the problem $\PLang[]{1wsat$^=$}$ below).
We present this connection in more generality than strictly necessary to prove
the upper bound since we will rely on variants of it later on. For propositional
formulas~$\psi$ in \emph{$d\Lang{cnf}$} (meaning \emph{at most} $d$ literals per
clause), let $\operatorname{vars}(\psi)$ and $\operatorname{clauses}(\psi)$
denote the sets of variables and clauses, respectively. For an \emph{assignment}
$\beta\colon\operatorname{vars}(\psi)\to\{0,1\}$, with the model relation $\beta
\models \psi$ defined as usual, the \emph{weight} is
$\operatorname{weight}(\beta)=\bigl|\bigl\{v\in\operatorname{vars}(\psi) \mid
\beta(v)=1\bigr\}\bigr|$. The following problem is the weighted version of the
satisfiability problem for $d\Lang{cnf}$ formulas:

\begin{problem}[{$\PLang[]{$d$wsat$^=$}$ for fixed $d$}]
\begin{parameterizedproblem}
  \instance A $d\Lang{cnf}$ formula~$\psi$ and a non-negative integer $k\in\mathbb{N}$.
  \parameter $k$
  \question Is there an assignment $\beta$ with $\beta\models\psi$ and
  $\operatorname{weight}(\beta) = k$? 
\end{parameterizedproblem}
\end{problem}
The problem is also known as $\PLang[]{$d$wsat}$ in the literature,
but we keep the ``$^=$'' superscript since we also consider
$\PLang[]{$d$wsat$^\le$}$ and $\PLang[]{$d$wsat$^\ge$}$, where we ask
whether there is a satisfying assignment with $\operatorname{weight}(\beta) \le 
k$ and $\operatorname{weight}(\beta) \ge k$, respectively. 
For us, the importance of these problems lies in the following lemma,
where ``$\le_{\Para\Class{AC}^0}$'' refers to the earlier-mentioned 
$\Para\Class{AC}^0$-reductions. Recall that these reductions
are extremely weak and that all classes considered in this
paper, including $\Para\Class{AC}^0$, are closed under them.

\begin{lemma~}\label{lemma:eso-to-sat}
  Let $d\ge 1$. Then:
  \begin{enumerate}
  \item For every $Q \in \PClass{FD}(E_1^{=}e^*a^d)$ we have $Q
    \le_{\Para\Class{AC}^0} \PLang[]{$d$wsat$^=$}$. 
  \item For every $Q \in \PClass{FD}(E_1^{\ge}e^*a^d)$ we have $Q
    \le_{\Para\Class{AC}^0} \PLang[]{$d$wsat$^\ge$}$. 
  \item For every $Q \in \PClass{FD}(E_1^{\le}e^*a^d)$ we have $Q
    \le_{\Para\Class{AC}^0} \PLang[]{$d$wsat$^\le$}$.
  \end{enumerate}
\end{lemma~}

\begin{proof~}
  In all three cases, $Q$ is the set of models of a weighted
  \textsc{eso} formula of the form $\exists^= X\, \phi(X)$ or
  $\exists^\le X\, \phi(X)$ or $\exists^\ge X\, \phi(X)$ where
  $\phi(X)$ has the quantifier pattern $e^*a^d$.
  In \cite[Lemma 7.2]{FlumG06} it is shown that given a formula
  $\phi(X)$ with such a pattern, we can map any structure
  $\mathcal S$ with some universe~$S$ to a $d\Lang{cnf}$
  formula~$\psi$ such that there there is a one-to-one correspondence
  between the sets $C \subseteq S$ with $\mathcal S \models \phi(C)$
  and the satisfying assignments~$\beta$ of~$\psi$. Furthermore, when
  $C$ corresponds to $\beta$, we have $|C| =
  \operatorname{weight}(\beta)$. While in \cite{FlumG06} it is only
  argued that the mapping from $\phi(X)$ to $\psi$ can be done in
  polynomial time, a closer look reveals that a $\Para\Class{AC}^0$ reduction
  suffices. This means that in all three items we can use this mapping
  as the reduction whose existence in claimed.
\end{proof~}

\begin{corollary~}\label{corollary:e-star-a}
  $\PClass{FD}(E_1^{=}e^*a)$, 
  $\PClass{FD}(E_1^{\le}e^*a)$, 
  $\PClass{FD}(E_1^{\ge}e^*a)$ are subsets of $\Para\Class{AC}^{0}$.
\end{corollary~}

\begin{proof~}
  Let us start with some $Q \in \PClass{FD}(E_1^{=}e^*a)$. By item~1 of
  Lemma~\ref{lemma:eso-to-sat}, $Q \le_{\Para\Class{AC}^0}
  \PLang[]{1wsat$^=$}$. Thus, showing $\PLang[]{1wsat$^=$} \in
  \Para\Class{AC}^0$ yields the claim for $\PClass{FD}(E_1^{=}e^*a)$ as
  $\Para\Class{AC}^0$  is closed under  
  $\Para\Class{AC}^0$ reductions. However, a $\Lang{1cnf}$ formula $\psi$ is
  just a   conjunction of literals. It is trivial to check in plain
  $\Class{AC}^0$ (independently of the parameter) whether $\psi$ is
  satisfiable (it may not contain a literal and its negation) and, if
  so, it is trivial to determine the single satisfying assignment
  $\beta \colon \operatorname{vars}(\psi) \to \{0,1\}$. We are left
  with having to check whether $\operatorname{weight}(\beta) = k$
  holds. It is well known~\cite{BannachST15} that the problem of checking
  whether the number of $1$ bits in a bitstring is at least, at most,
  or equal to a parameter value lies in $\Para\Class{AC}^0$, yielding
  the claim. However, this also yields the other two items.
\end{proof~}

\subsection{Solution Weight Is At Least the Parameter for Arbitrary
  Structures}
\label{section:arb-max}

For $\PClass{FD}(E_1^{\ge}p)$, we get the exact same dichotomy as for
$\PClass{FD}(E_1^{=}p)$. However, a look at the detailed bounds in the
lemma shows that for $\PClass{FD}(E_1^{\ge}p)$ we only get a lower bound
for \emph{undirected} graphs and not for \emph{basic} graphs (and,
indeed, we will show in Section~\ref{section:weighted-fd-basic} that
the complexity is different for basic graphs). Furthermore, while we
always have $\PClass{FD}(E_1^= p) \subseteq \Class W[t]$ for some~$t$ (see
\cite[Definition 5.1]{FlumG06}), we show that the patterns $E_1^{\ge}eae$ or
$E_1^{\ge}aee$ suffice to describe even $\Para\Class{NP}$-complete
problems even on basic graphs. Thus, although the tractability
frontier (``in $\Class{FPT}$'' versus ``contains $\Class{W}[1]$-hard
problems'') is the same for $\PClass{FD}(E_1^{=}p)$ and $\PClass{FD}(E_1^{\ge}p)$, 
the detailed structure is more complex.

\begin{theorem}[Complexity Dichotomy for {$\PClass{FD}(E_1^{\ge}p)$}]\label{theorem:arb-max}
  Let $p$ be a pattern.
  \begin{enumerate}
  \item $\PClass{FD}(E_1^{\ge}p) \subseteq \Para\Class{AC}^{0}$, if $p
    \preceq e^*a$.
  \item $\PClass{FD}(E_1^{\ge}p)$ contains a $\Class{W[1]}$-hard
    problem, if $aa\preceq p$ or $ae\preceq p$.
  \end{enumerate}
  Both items also hold for $\PClass{FD}_{\mathrm{undirected}}(E_1^{\ge}p)$.
\end{theorem}
The theorem follows directly from the following lemma (whose last two
items are not actually needed here, but shed more
light on the detailed structure and \emph{will} be needed in
Section~\ref{section:weighted-fd-basic}).

\begin{lemma~}[Detailed Bounds for {$\PClass{FD}(E_1^{\ge}p)$}]\label{lemma:arb-max}
  \hfil
  \begin{enumerate}
  \item\label{item:max-e*a} $\PClass{FD}(E_1^{\ge}e^*a) \subseteq
    \Para\Class{AC}^{0}$.
  \item\label{item:max-aa} $\PClass{FD}_{\mathrm{basic}}(E_1^{\ge}aa)$ contains a $\Class{W[1]}$-hard problem.
  \item\label{item:max-ae} $\PClass{FD}_{\mathrm{undirected}}(E_1^{\ge}ae)$ contains a $\Para\Class{NP}$-hard problem.
  \item\label{item:max-eae} $\PClass{FD}_{\mathrm{basic}}(E_1^{\ge}eae)$ contains a $\Para\Class{NP}$-hard problem.
  \item\label{item:max-aee} $\PClass{FD}_{\mathrm{basic}}(E_1^{\ge}aee)$ contains a $\Para\Class{NP}$-hard problem.
  \end{enumerate}
\end{lemma~}
\begin{proof~}
  Item~\ref{item:max-e*a} is already stated in Corollary~\ref{corollary:e-star-a}. For
  item~\ref{item:max-aa}, equation~\ref{eq:clique} shows that the
  $\Class{W}[1]$-complete problem $\PLang{clique}$ can be expressed
  with a weighted \textsc{eso} formula with the pattern
  $E^{\ge}_1aa$ and, thus, lies in
  $\PClass{FD}_{\mathrm{basic}}(E_1^{\ge}aa)$.

  For the other items, a claim is useful:
  \begin{claim}\label{claim:np-to-para-np} 
    If there is an $\Class{NP}$-hard problem
    in $\Class{FD}(E_1p)$, there is a $\Para\Class{NP}$-hard problem in
    $\PClass{FD}(E_1^{\ge}p)$; and this holds also for the restrictions to
    undirected, basic, or directed graphs. 
  \end{claim}
  To see that this claim holds, just note that the non-parameterized
  problem is the special case of the parameterized maximization
  problem where $k = 0$.

  To prove item~\ref{item:max-ae}, observe that Gottlob et al.\ have
  shown \cite[Theorem 2.1]{GottlobKS04} that there are
  $\Class{NP}$-complete problems in $\Class{FD}_{\mathrm{undirected}}(E_1ae)$. 
  By the claim, there must be $\Para\Class{NP}$-hard problems in
  $\PClass{FD}_{\mathrm{undirected}}(E_1^{\ge}ae)$. Next, for item~\ref{item:max-eae},
  in~\cite[Theorem 2.5]{GottlobKS04} it is shown that there is 
  an $\Class{NP}$-hard problem in
  $\Class{FD}_{\mathrm{basic}}(E_1eae)$. Finally, for
  item~\ref{item:max-aee}, by~\cite[Theorem 2.6]{GottlobKS04} there is
  also an $\Class{NP}$-hard problem in
  $\Class{FD}_{\mathrm{basic}}(E_1aee)$. 
\end{proof~}

\noindent Note that compared to Lemma~\ref{lemma:arb-equal}, in the third
item of Lemma~\ref{lemma:arb-max} we have shown the lower
bound for the restriction of structures to undirected graphs rather
than basic graphs. This is no coincidence: the self-loops which are
allowed for undirected graphs have in some cases an impact on the
complexity of the problems we can express. Later, when we cover basic
graphs, we will see that, indeed, sometimes problems become easier
compared to their counterparts where undirected graphs are admissible as
structures.

\subsection{Solution Weight Is At Most the Parameter for Arbitrary Structures}

When the parameter is an upper bound on the weight of solutions, the
tractability landscape changes quite a bit: The pattern $E_1^\le ae$
becomes intractable, while $E_1^\le aa$ no longer lies in
$\Para\Class{AC}^0$, but stays tractable:

\begin{theorem}[Complexity Trichotomy for {$\PClass{FD}(E_1^{\le}p)$}]\label{theorem:arb-min}
  Let $p$ be a pattern.
  \begin{enumerate}
  \item $\PClass{FD}(E_1^{\le}p) \subseteq \Para\Class{AC}^{0}$, if $p
    \preceq e^*a$.
  \item $\PClass{FD}(E_1^{\le}p) \subseteq \Para\Class{AC}^{0\uparrow}$
    but  $\PClass{FD}(E_1^{\le}p) \not\subseteq \Para\Class{AC}^{0}$, if
    $aa \preceq p \preceq e^*a^*$.
  \item $\PClass{FD}(E_1^{\le}p)$ contains a $\Class{W[1]}$-hard
    problem, if $ae\preceq p$.
  \end{enumerate}
  All items also hold for $\PClass{FD}_{\mathrm{undirected}}(E_1^{\le}p)$.
\end{theorem}
As before, the theorem covers all possible~$p$ and follows from a
lemma that is a bit more general than strictly necessary: We will need items~\ref{item:min-aaa} and
\ref{item:min-eaa} only in Section~\ref{section:aa}, where we show
that item~\ref{item:min-aa} does \emph{not} hold for basic graphs.

\begin{lemma}[Detailed Bounds for {$\PClass{FD}(E_1^{\le}p)$}]\label{lemma:arb-min}
  \hfil
  \begin{enumerate}
  \item\label{item:min-e*a} $\PClass{FD}(E_1^{\le}e^*a) \subseteq
    \Para\Class{AC}^{0}$.
  \item\label{item:min-e*a*}  $\PClass{FD}(E_1^{\le}e^* a^*) \subseteq
    \Para\Class{AC}^{0\uparrow}$.
  \item\label{item:min-aa}  $\PClass{FD}_{\mathrm{undirected}}(E_1^{\le}aa)$ contains a problem not in $\Para\Class{AC}^{0}$.
  \item\label{item:min-aaa}  $\PClass{FD}_{\mathrm{basic}}(E_1^{\le}aaa)$ contains a problem not in $\Para\Class{AC}^{0}$.
  \item\label{item:min-eaa}  $\PClass{FD}_{\mathrm{basic}}(E_1^{\le}eaa)$ contains a problem not in $\Para\Class{AC}^{0}$.
  \item\label{item:min-ae}  $\PClass{FD}_{\mathrm{basic}}(E_1^{\le}ae)$ contains a $\Class{W[2]}$-hard problem.
  \end{enumerate}
\end{lemma}
\begin{proof}
  Item~\ref{item:min-e*a} is already stated in
  Corollary~\ref{corollary:e-star-a}. Item~\ref{item:min-e*a*} is shown in
  Lemma~\ref{lemma:arb-min-2} below. Items~\ref{item:min-aa},
  \ref{item:min-aaa}, and \ref{item:min-eaa} are shown in
  Lemma~\ref{lemma:arb-min-3}, Lemma~\ref{lemma:basic-min-1}, and
  Lemma~\ref{lemma:basic-min-2}, respectively.
  Item~\ref{item:min-ae} follows, once more, from
  $\PLang{dominating-set} \in
  \PClass{FD}_{\mathrm{basic}}(E_1^{\le}ae)$ by equation~\eqref{eq:domset}.
\end{proof}

\begin{lemma~}\label{lemma:arb-min-2}
  $\PClass{FD}(E_1^{\le}e^* a^*) \subseteq \Para\Class{AC}^{0\uparrow}$.
\end{lemma~}
\begin{proof~}
  Let $Q \in \PClass{FD}(E_1^{\le}e^* a^d)$ for some fixed~$d$. By
  Lemma~\ref{lemma:eso-to-sat}, $Q \le_{\Para\Class{AC}^0}
  \PLang{$d$wsat$^{\le}$}$. We now show that $\PLang{$d$wsat$^{\le}$} \in
  \Para\Class{AC}^{0\uparrow}$, which implies the claim as
  $\Para\Class{AC}^{0\uparrow}$ is closed under $\Para\Class{AC}^0$
  reductions.

  We have to construct a circuit family of depth $f(k)$ and size
  $f(k)\cdot n^{O(1)}$ for $n= \left|\operatorname{vars}(\psi)\right|$ for some
  computable function~$f$. The circuit implements a bounded
  search tree such that every layer evaluates one level of the
  tree. To that end, each layer gets a set $\Psi_i$ of formulas as
  input and outputs a new set $\Psi_{i+1}$ of formulas. We start with
  $\Psi_0 = \{\psi\}$. The invariant will be that $\psi$ has a
  satisfying assignment of (exact) weight $w$ iff some formula in $\Psi_i$ has
  a satisfying assignment of (exact) weight $w-i$. 

  To compute the next $\Psi_{i+1}$ for $i \in \{0,\dots,k\}$, we
  perform the following operations in parallel for every $\rho \in \Psi$:
  \begin{enumerate}
  \item If every clause in $\rho$ contains a negative literal (meaning
    that $\rho$ is satisfied by the all-$0$ assignment), accept the
    original input. (Doing so is correct by the invariant.)
  \item Take a clause $c\in\operatorname{clauses}(\rho)$ that only contains
    positive literals $x_1$, $\dots$, $x_e$. For each $x_i$, generate
    a new formula $\rho^i$ from $\rho$ by ``setting one of these
    variables  to~$1$'' or, formally, by removing all clauses that
    contain it positively and removing the variable from all clauses
    that contain it negatively, respectively. Add $\rho^1$ to $\rho^e$
    to $\Psi_{i+1}$. (This maintains the invariant as we \emph{must}
    set one of the $x_i$ to $1$ in any assignment that satisfies~$\rho$.)
  \end{enumerate}
  If we have not accepted the input after having computed
  $\Psi_{k+1}$, we reject. This is correct since all satisfying
  assignments of the $\rho \in \Psi_{k+1}$ have weight at least~$0$
  and, thus, by the invariant all satisfying assignments of $\psi$ have
  weight at least $k+1 - 0 > k$.

  To see that the circuit can be implemented with the claimed depth
  and size, note that since $e \le d$, the list grows by a factor of
  at most $d$ in every layer and we can implement each layer in
  constant depth. As there are only $k+1$ layers, we have
  $|\Psi_{k+1}| \le (k+1)^d \eqcolon f(k)$. 
\end{proof~}
For the three remaining still-to-be-proved lower bounds in
Lemma~\ref{lemma:arb-min}, the claim is always that a class is
(unconditionally) not contained in  
$\Para\Class{AC}^0$. To prove this, we will show that the following
problem is (provably) not in $\Para\Class{AC}^0$ but can be
$\Para\Class{AC}^0$-reduced to problems that lie in the three
classes:

\begin{problem}[{$\PLang[]{matched-reach}$}]
  \begin{parameterizedproblem}
    \instance A directed layered graph~$G$ with vertex set
    $\{1,\dots,n\} \times \{1,\dots,k\}$, where the $i$th \emph{layer}
    is $V_i \coloneq
    \{1,\dots,n\} \times \{i\}$, such that
    for each $i\in\{1,\dots,k-1\}$ the edges point to the next layer
    and they form a perfect matching between $V_i$ and $V_{i+1}$;
    two designated vertices $s \in V_1$ and $t \in V_k$. 
    \parameter k.
    \question Is $t$ reachable from $s$?
  \end{parameterizedproblem}
\end{problem}

\begin{lemma~}\label{lemma:matched}
  $\PLang[]{matched-reach} \notin \Para\Class{AC}^0$ and consequently, for
  any problem $Q$ with $\PLang[]{matched-reach}
  \le_{\Para\Class{AC}^0} Q$ we have $Q \notin  \Para\Class{AC}^0$.
\end{lemma~}
\begin{proof~}
  Beame et al.~\cite{BeameIP98} have shown that any depth-$c$ circuit
  that solves $\PLang[]{matched-reach}$ requires size
  $n^{\Omega(k^{(\rho^{-2c})/3})}$, where $\rho$ is the golden 
  ratio. However, $\PLang[]{matched-reach} \in \Para\Class{AC}^0$
  would imply that for some 
  constant $c$ there is a depth-$c$ circuit family that decides
  the problem in size $f(k) \cdot n^{O(1)}$; contradicting the Beame
  et al.\ bound of $n^{k^{\Theta(1)}}$.
  For the claim concerning~$Q$, just note that $\Para\Class{AC}^0$ is
  closed under $\Para\Class{AC}^0$ reductions.
\end{proof~}

\begin{lemma~}\label{lemma:min-aa-not-in-ac0}\label{lemma:arb-min-3}
  $\PClass{FD}_{\mathrm{undirected}}(E_1^{\le}aa)\not\subseteq\Para\Class{AC}^{0}$.
\end{lemma~}

\begin{proof~}
  Consider the following formula with quantifier pattern $E^\le_1 aa$:
  \begin{align*}
    \phi_{\text{reach}} \coloneq \exists^\le S\,\forall x \forall y
    \bigl(\bigl((x \sim x) \to Sx\bigr) \land
    \bigl((Sx \land x \sim y) \to Sy\bigr)\bigr). 
  \end{align*}
  We claim that we can reduce $\PLang[]{matched-reach}$ to
  $\PLang{models}_{\mathrm{undirected}}(\phi_\text{reach})$ as
  follows (and the claim then follows from Lemma~\ref{lemma:matched}):
  On input $(G,s,t)$, the reduction first checks that the graph is,
  indeed, a layered graph with perfect matchings between consecutive
  levels. Then, we forget about the direction of the 
  edges (making the graph undirected). Next, we add an additional layer
  $V_{k+1}$ and match each vertex of $V_k$ to the corresponding
  new vertex~$V_{k+1}$. Next, we \emph{remove} the just-added edge
  from~$t$ in layer~$V_k$ to its counterpart in
  layer~$V_{k+1}$. Finally,  
  add a self-loop at~$s$. To see that this reduction is correct, note
  that the self-loop at $s$ forces it (but does not \emph{force} any
  other vertex), to be part of the
  solution set $S$ by the first part of the formula. The second part
  then forces the solution set to be closed under reachability. Thus,
  if $t$ lies on the same path as~$s$, there is a solution of size
  $k$, and if not, the smallest solution has size  $k + 1$. 
\end{proof~}

\begin{lemma~}\label{lemma:basic-min-1}
  $\PClass{FD}_{\mathrm{basic}}(E_1^{\le} aaa)\not\subseteq\Para\Class{AC}^{0}$.
\end{lemma~}
\begin{proof~}
  We reduce $\PLang{matched-reach}$ to
  $\PLang{models}_{\mathrm{basic}}(\phi_\text{reach-aaa})$ for 
  \begin{align*}
    \phi_{\text{reach-aaa}} \coloneq \exists^{\le} S\, \forall x\forall y\forall z \big(((x\sim y \land y\sim z \land x\sim z)\rightarrow Sx)\land ((Sx \land x \sim y) \rightarrow Sy)\big).
  \end{align*}
  On input $(G,s,t)$, once more we start by forgetting about the
  direction of the edges. This time, add two vertices and connect them
  to~$s$ so that these three vertices form a triangle. Do the same
  for~$t$ by adding another two vertices. Output $k+4$ as the new
  parameter. This reduction is  correct, because the first part of the
  formula forces every vertex which is part of a triangle to be part
  of~$S$, which are exactly the triangles at $s$ and~$t$. The latter
  part of the formula forces the solution set to be closed under
  reachability. Thus, if $t$ lies on the same path as~$s$, there is a
  solution of size $k + 4$, and if not, the smallest solution has size
  as least $k + 5$. 
\end{proof~}

\begin{lemma~}\label{lemma:basic-min-2}
  $\PClass{FD}_{\mathrm{basic}}(E_1^{\le} eaa)\not\subseteq\Para\Class{AC}^{0}$.
\end{lemma~}

\begin{proof~}
  We reduce $\PLang[]{matched-reach}$ to
  $\PLang{models}_{\mathrm{basic}}(\phi_\text{reach-eaa})$ for
  \begin{align*}
    \phi_{\text{reach-eaa}} \coloneq \exists^{\le} S\, \exists z\forall
    x\forall y \big(Sz \land ((Sx \land x \sim y) \rightarrow
    Sy)\big). 
  \end{align*}
  On input $(G,s,t)$ we forget the direction of edges and add a single
  vertex that we connect to every vertex that has 
  degree~$1$ except for $s$ and~$t$. If $t$ is on the same path
  as~$s$, there will be two connected components: One consisting of the
  path between $s$ and $t$, and one containing everything else. In
  particular, there is a component of size $k$ and one of size $n -
  k$. If $t$ is not on the same path as~$s$, there is just a single
  connected component of size~$n$. To see that the reduction is
  correct, notice that the first part of the formula ($Sz$) 
  forces at least one vertex to be part of the solution. The latter part of the
  formula once more forces the solution set to be closed under reachability. By
  construction, there is a solution of size at most $k$ iff $t$ was on the same
  path as~$s$.
\end{proof~}

%
%

\section{Classifying Parameterized ESO Classes: Basic Graphs}
\label{section:weighted-fd-basic}

We saw in Section~\ref{section:weighted-fd} that the parameterized complexity of
weighted \textsc{eso} classes depends strongly on the first-order
quantifier pattern~$p$ and on whether we are interested in the
equal-to, at-least, or at-most case -- but it does \emph{not} matter
whether we consider arbitrary logical structures, only directed
graphs, or only undirected graphs: the complexity is always the
same.
The situation changes if we restrict attention to \emph{basic graphs,}
which are undirected graphs without self-loops: We get different
tractability frontiers. This is an interesting effect since the only
difference 
between undirected graphs and basic graphs is that some vertices may
have self-loops -- and self-loops are usually neither needed nor used
in hardness proofs, just think of the clique problem, the vertex cover
problem, or the dominating set problem. Nevertheless, it turns out
that ``a single extra bit per vertex'' and sometimes even ``a single
self-loop'' allows us to encode harder problems than without.

To establish the tractability frontier for basic graphs, we can, of
course, recycle many results from the previous section: Having a look
at the detailed bounds listed in Lemmas~\ref{lemma:arb-equal},
\ref{lemma:arb-max}, and \ref{lemma:arb-min}, we see that the upper 
bounds are shown for arbitrary structures and, hence, also hold for
basic graphs; and many lower bounds have also already been established
for basic graphs. Indeed, it turns out there are exactly two classes
whose complexity ``changes'' when we restrict the inputs to basic
graphs: 
\begin{enumerate}
\item $\PClass{FD}_{\mathrm{basic}}(E_1^\ge ae)$ lies in $\Para\Class{AC}^0$,
  while $\PClass{FD}(E_1^\ge ae)$ does not.
\item $\PClass{FD}_{\mathrm{basic}}(E_1^\le aa)$ lies in $\Para\Class{AC}^0$,
  while $\PClass{FD}(E_1^\le aa)$ does not.
\end{enumerate}
We have already shown the ``while \dots'' part in
Section~\ref{section:weighted-fd}, it is the upper bounds that are new. For all other
patterns~$p$, the classification does not change.
Proving the two items turns out to be technical and we
devote one subsection to each of these results.

%
%
%
%
\subsection[The Case Greater-Than Case for ae and Basic Graphs]
{The Case \textit{E$_{\text{\upshape 1}}^{\,\ge}$ae} for Basic Graphs}

\label{section:ae}

As mentioned, for the classification of the complexity of
$\PClass{FD}_{\mathrm{basic}}(E_1^\ge p)$ we can reuse all of our
previous results, \emph{except} that
$\PClass{FD}_{\mathrm{basic}}(E^\ge_1 ae) \subseteq \Para\Class{AC}^0$
holds. This is the statement of Lemma~\ref{lemma:basic-ae}, which is
proved in the rest of this section. However, before be plunge into the
glorious details, let us 
ascertain that there are no further patterns $p \neq ae$ for which
$\PClass{FD}_{\mathrm{basic}}(E^\ge_1 p)$ becomes any easier: To see
this, note that for any $p$ with $p \not\preceq ae$ we have $aa
\preceq p$ or $eae \preceq p$ or $aee \preceq p$; and for $aa$, $eae$,
and $aee$ we have already established hardness for \emph{basic} graphs
in Lemma~\ref{lemma:arb-max}. For completeness, we spell out the
resulting structure:

\begin{theorem}[Dichotomy for {$\PClass{FD}_{\mathrm{basic}}(E_1^{\ge}p)$}]\label{theorem:basic-max}
  Let $p$ be a pattern.
  \begin{enumerate}
  \item $\PClass{FD}_{\mathrm{basic}}(E_1^{\ge}p) \subseteq \Para\Class{AC}^{0}$, if $p
    \preceq e^*a$ or $p \preceq ae$.
  \item $\PClass{FD}_{\mathrm{basic}}(E_1^{\ge}p)$ contains a $\Class{W[1]}$-hard
    problem, if $aa\preceq p$, $eae\preceq p$, or $aee\preceq p$.
  \end{enumerate}
\end{theorem}

\begin{lemma}[\iftcsautomove$\blacktriangledown$\fi]\label{lemma:basic-ae}
  $\PClass{FD}_{\mathrm{basic}}(E^\ge_1 ae) \subseteq
  \Para\Class{AC}^0$.
\end{lemma}
For the surprisingly difficult proof we employ machinery first
used in~\cite{GottlobKS04} and in~\cite[Section 3.3]{Tantau15}: Our
objective is to represent the problems in 
$\PClass{FD}_{\mathrm{basic}}(E_1^{\ge}ae)$ as special kinds of graph
coloring problems -- and to then show that we can solve these problem
in $\Para\Class{AC}^0$. \iftcsautomove The (very) technical details can be
found on pages~\pageref{page:lemma-ae-first} to
\pageref{page:lemma-ae-last} in the appendix.\fi

\tcsautomoveaddto{main}{
  \subsection{Proofs for Section~\ref{section:ae}}

  There is just one lemma in the section whose proof is still missing:
  Lemma~\ref{lemma:basic-ae} (which claimed
  $\PClass{FD}_{\mathrm{basic}}(E_1^{\ge}ae) \subseteq
  \Para\Class{AC}^0$). However, proving this lemma necessitates 
  a \emph{lot} of additional technical machinery, so let us start with
  some definitions.\label{page:lemma-ae-first}

}

\begin{scope~}

\subparagraph*{Definition of Pattern Graphs and Saturated Basic
  Graphs.}
Following \cite{GottlobKS04}, a \emph{pattern graph} $P =
(C, A^\oplus, A^\ominus)$ consists of a set of \emph{colors} $C$, a
set $A^\oplus \subseteq C\times C$ of $\oplus$-arcs, and a set
$A^\ominus\subseteq C\times C$ of $\ominus$-arcs (note that $A^\oplus$
and $A^\ominus$ need not be disjoint). In our paper, we will only need
the case that there are only two colors, so $C = \{\text{black},
\text{white}\}$ will always hold, and we call such a pattern graph
\emph{binary}. In the rest of the section, \emph{pattern} always
refers to a \emph{binary pattern graph} (and no longer to quantifier
prenex patterns -- we are only interested in the single pattern
$E_1^\ge ae$ anyway). Observe that there 256 possible binary pattern
graphs. A \emph{superpattern} of a pattern $P = (C, A^\oplus,
A^\ominus)$ is any pattern $P' = (C, B^\oplus, B^\ominus)$ with
$A^\oplus \subseteq B^\oplus$ and $A^\ominus \subseteq B^\ominus$. A
\emph{$\oplus$-superpattern} is a superpattern with $A^\ominus =
B^\ominus$.

For a basic graph $B = (V, E)$, a \emph{coloring
of~$B$} is a function $c\colon V\to C$. However, unlike standard
coloring problems, where vertices connected by an edge must have
different colors, what constitutes an allowed coloring is dictated by
the pattern graph via a \emph{witness function:} A mapping
$w\colon V\to V$ is called a \emph{witness function for a coloring
$c$} if for all $x\in V$ we have 
\begin{enumerate}
\item $x\neq w(x)$,
\item if $\{x, w(x)\}\in E$, then $\bigl(c(x), c(w(x))\bigr) \in A^\oplus$, and
\item if $\{x, w(x)\}\not\in E$, then $\bigl(c(x), c(w(x))\bigr)\in A^\ominus$.
\end{enumerate}
The idea is that a vertex $x$ and its witness $w(x)$ are connected by
``a $\oplus$-arc'' if there is an edge between them and by ``a
$\ominus$-arc'' if there is no edge between them. The pattern graph
then tells us which colors are allowed for $x$ and $w(x)$ in
dependence on which kind of arc there is. For instance, for the
pattern \patterntwo{any,draw=none,overlay}{}{any,draw=none,overlay}{}{->}{$\oplus$}{->}{$\oplus$}
every vertex must be connected by an edge to a vertex of the opposite
color. Note that this is not the same as asking for a 2-coloring:
We only impose a requirement on the edge (corresponding to a
$\oplus$-arc) between
$x$ and $w(x)$, other edges are not relevant. In more detail, consider
a triangle with the vertices $\{x,y,z\}$ and the coloring $c(x) =
\mathrm{black}$, $c(y) = c(z) = \mathrm{white}$ and the witness
function $w(x) = y$ and $w(y) = w(z) = x$. Then the coloring is legal
with respect to the pattern and the witness function, despite that
fact that a triangle is not 2-colorable.

If there exists a coloring~$c$ together with a witness function~$w$
for $B$ with respect to~$P$, we say that \emph{$B$ is $P$-saturated by
$c$ and~$w$}. The saturation problem $\Lang{saturation}(P)$ for a
pattern~$P$ is then simply the set of all basic graphs $B =
(V,E)$ that can be $P$-saturated (via some coloring $c$ and witness
function~$w$). 
The relation between the saturation problem and $E_1 ae$ is as
follows: We want the witness function to tell us for 
each~$x$ in $\forall x$ which~$y$ in $\exists y$ we must pick to make a formula
of the form $\exists S\, \forall x\exists y\; \psi$ true: \emph{We color a
vertex black to indicate that it should be included in~$S$, otherwise
we color it white.} In this way, one can associate a pattern graph with
each $E_1ae$-formula.

\begin{fact}[{\cite[Fact 3.3]{Tantau15} for a single quantifier}]\label{fact:fd-to-satu}
  For every \textsc{eso} formula $\phi$ with quantifier pattern
  $E_1ae$ there is a binary pattern graph~$P$ such that
  $\Lang{models}(\phi) = \Lang{saturation}(P)$.  
\end{fact}
(Strictly speaking, this only holds for basic graphs $B$ with at least
two vertices. For this reason, in the following we always assume that
$|V| \ge 2$ holds.) 

Adapting this approach to the parameterized setting is
straightforward: Define the \emph{weight} of a binary coloring as the
number of vertices that are colored black. This leads to the following 
parameterized problem and transfer of Fact~\ref{fact:fd-to-satu} to
the parameterized setting: 

\begin{problem}[{$\PLang{saturation}^\ge(P)$ for a fixed binary pattern
      graph $P = (C, A^\oplus, A^\ominus)$}]
\begin{parameterizedproblem}
  \instance A basic graph $B = (V, E)$ and an integer $k\in \mathbb{N}$.
  \parameter $k$.
  \question Can $B$ be $P$-saturated via a coloring of weight
  \emph{at least}~$k$?
\end{parameterizedproblem}
\end{problem}

\begin{lemma}\label{lemma:fd-to-satu}
  For every weighted \textsc{eso} formula $\phi$ with quantifier pattern
  $E_1^\ge ae$ there is a binary pattern graph~$P$ such that
  $\PLang{models}(\phi) = \PLang{saturation}^\ge(P)$.  
\end{lemma}

\begin{proof}
  A detailed proof of Fact~\ref{fact:fd-to-satu} can be found in the
  paper by Gottlob, Kolaitis, and Schwentick~\cite[Theorem
    4.6]{GottlobKS04}, we just sketch the general idea and how
  parameterization enters the picture. In the proof of the fact,
  multiple existential second-order quantifiers are considered; we
  only have a single one and in result just get two colors.

  The idea is that, in the first direction, assume that a graph~$B$ is
  a model of the formula $\phi = \exists^\ge S\, \forall x\forall
  y\;\psi$. Then any size-$s$ interpretation of the set variable~$S$
  that makes the formula true gives rise to a weight-$s$ coloring~$c$
  that simply colors all vertices in~$S$ black and all other vertices
  white. Next, the formula states that for all interpretations of~$x$,
  which correspond to the vertices~$v$ of the input graph~$B$, there is an 
  interpretation of~$y$, corresponding to the
  witness~$w(v)$. The quantifier-free part $\psi$ now imposes certain
  conditions on what colors $v$ and $w(v)$ may have in dependence on
  whether or not there is an edge between them or not. These
  dependencies can then be encoded into the presence or absence of
  arcs in $A^\oplus$ and $A^\ominus$, see \cite[Theorem
    4.6]{GottlobKS04} for details. Note that, clearly, by construction,
  the weight of the solutions~$S$ and of the coloring~$c$ are the
  same. 

  For the other direction, if a graph~$B$ can be saturated by~$P$, the
  coloring indicates (in the form of the black vertices) which
  elements should be contained in the 
  interpretation of~$S$, and the witness function $w$ indicates 
  how $y$ should be interpreted for every interpretation of~$x$ to
  make the formula true. Once more the weights are the same.
\end{proof}

\subparagraph*{General Tools for Pattern Graphs.}
With these preparations, our ultimate goal of proving
Lemma~\ref{lemma:basic-ae} (which claimed
$\PClass{FD}_{\mathrm{basic}}(E_1^{\ge}ae) \subseteq
\Para\Class{AC}^0$) is achieved if 
we can show that $\PLang{saturation}^\ge(P) \subseteq \Para\Class{AC}^0$
holds for all (binary) patterns~$P$. Before we go over the different cases, we
develop some simple tools that will prove useful repeatedly in the
different cases.

We start with some observations concerning which patterns we
need to consider. First, a simple, but useful observation
is the following, which will often allow us to considerably reduce the
number of cases we need  to consider:
\begin{lemma}\label{lemma:symmetric}
  For $P = (C, A^\oplus, A^\ominus)$ let $\bar P = (C, A^\ominus,
  A^\oplus)$. Then $\PLang{saturation}^\ge(P) \in \Para\Class{AC}^0$
  iff $\PLang{saturation}^\ge(\bar P) \in \Para\Class{AC}^0$.
\end{lemma}
\begin{proof}
  Any $B$ can be $P$-saturated iff $\bar B$ can be $\bar P$-saturated
  (by the same coloring and same witness function), where $\bar B$
  results from $B$ by exchanging whether there is an edge or not for
  any two element vertex set $\{x,y\}$ (so we exchange edges and
  non-edges, but do not  add self-loops).
\end{proof}

Another observation concerning the patterns is the following:
\begin{lemma}\label{lemma:onlycycles}
  Let $P'$ result from $P = (C, A^\oplus, A^\ominus)$ by removing an
  arc $(c_1,c_2)$ from $A^\oplus$ or from $A^\ominus$ such that no arc
  in $A^\oplus \cup A^\ominus$ starts at~$c_2$. Then
  $\PLang{saturation}^\ge(P) = \PLang{saturation}^\ge(P')$.
\end{lemma}
\begin{proof}
  The arc can never be used by any witness function for any basic
  graph~$B$: Suppose that $B$ is $P$-saturated by a coloring~$c$ and a 
  witness function~$w$ such that for some vertex $v$ we have $c(v) =
  c_1$ and $c(w(v)) = c_2$. Consider $u \coloneq w(w(v))$. Since no arc
  leaves $c_2$, no matter what color we assign to~$u$, the coloring
  will not be legal. Thus, the arc $(c(u),c(v))$ is never used and, hence,
  $c$ and $w$ also show that $B$ can be $P'$-saturated (with the same
  weight). 
\end{proof}

The next ideas concern algorithmic tools for showing membership in
$\Para\Class{AC}^0$. In several cases, we will use \emph{reduction
rules}. Such a rule takes a pair $(B,k)$ as input and outputs a new
pair $(B',k')$ with smaller $B'$ and possibly smaller~$k'$ such that
the pairs are membership equivalent with respect to
$\PLang{saturation}^\ge(P)$. The objective is to arrive at some
$B'=(V',E')$ for which we can decide membership in 
$\PLang{saturation}^\ge(P)$ in $\Para\Class{AC}^0$ -- and, thereby, for
the original input. Note that unlike standard kernelization
methods in parameterized complexity, we need to make sure that the
application of the reduction rules can be done in $\Class{AC}^0$ or at least
in $\Para\Class{AC}^0$ and, crucially, if we apply several reduction
rules in sequence, this sequence needs to have a \emph{constant
length} that is independent of $B$ and~$k$ (so that when we
implement these rules as circuit layers of constant depth, the total
depth is still constant).

Two cases will be especially useful as
``endpoints'' for applications of reduction rules: 
\begin{lemma}\label{lemma:brute}
  There is a $\Para\Class{AC}^0$ circuit that on input $(B,k)$
  correctly decides membership in 
  $\PLang{saturation}^\ge(P)$ whenever $|V| \le f(k)$ for some fixed
  computable function~$f$.
\end{lemma}
\begin{proof}
  Use brute force: Hardwire all members of $\PLang{saturation}^\ge(P)$
  of size at most $f(k)$ into a constant-depth circuit.
\end{proof}
\begin{lemma}\label{lemma:slice}
  There is a $\Para\Class{AC}^0$ circuit that on input $(B,0)$
  correctly decides membership in $\PLang{saturation}^\ge(P)$.
\end{lemma}
\begin{proof}
  The case $k=0$ is just the $E_1ae$ case, for which
  \cite[Section~3.3]{Tantau15} shows membership in $\Class{FO} =
  \Class{AC}^0$.
\end{proof}

\begin{remark*}
  We are glossing over some encoding issues concerning reduction rules
  and the above two lemmas (whereas such issues are not important in
  the rest of this paper, one has to be a bit careful here): In the
  following, several reduction rules will be of the form ``If \dots\
  holds, remove all vertices and adjacent edges in some set $U$
  from~$B$ to produce a new graph~$B'$.'' The problem is that when $B$
  is given in a standard encoding like an adjacency matrix, if $|U|$
  is unbounded (in terms of~$k$) and the members of $U$ are
  ``scattered around'' in the set~$V$ of~$B$'s vertices, we have no
  way of computing the adjacency matrix of~$B'$ using a
  $\Para\Class{AC}^0$ circuit: We would need to repeatedly ``count how
  many elements of~$U$ come before some position~$i$ in the encoding
  of~$B$''. For this reason, we allow basic graphs $B = (V,E)$ to be
  encoded using adjacency matrices of possibly larger basic graphs
  $B^+ = (V^+, E^+)$ with $V \subseteq V^+$ and $E \subseteq E^+$
  together with a bitstring of length $|V^+|$ that is set to~$1$ for
  all $v \in V^+$ with $v \in V$ and set to~$0$ otherwise. With this
  encoding, ``removing vertices from~$B$'' then just means setting
  some bits in the bitstring to~$1$. Crucially, this encoding still
  allows us to answer questions like ``Are there $2k$ many neighbors of
  some vertex $v$ in~$B$?'' using $\Para\Class{AC}^0$ circuits since we
  can check parameter-dependent thresholds and can trivially restrict
  the counting to the elements for which the bitstring indicates
  membership in~$V$.
\end{remark*}

For some patterns~$P$, it turns out that
``sufficiently large'' basic graphs~$B$ can \emph{always} be
$P$-saturated and, even better, can \emph{always be $P$-saturated by a
weight-$k$-or-more coloring.} In such cases, we can  accept input
graphs $B$ ``just because they are large'' -- and if they are small,
Lemma~\ref{lemma:brute} applies. Since we will use this argument quite
often, let us say that $P$ is \emph{heavily saturating on large graphs
(after preprocessing)} if there are a computable function~$f$ and
(optionally) a $\Para\Class{AC}^0$-computable reduction rule (in the
sense described earlier) that maps any $(B,k)$ to some $(B',k')$ such
that whenever $B'$ has at least $f(k')$ many vertices, then $B'$ can be
$P$-saturated via a coloring of weight at least~$k'$.

\begin{lemma}\label{lemma:heavy}
  If $P$ is heavily saturating on large graphs, then
  $\PLang{saturation}^\ge(P) \in \Para\Class{AC}^0$. 
\end{lemma}

\begin{proof}
  On input $(B,k)$ use the reduction rule to compute $(B',k')$. Let
  $B'= (V',E')$. If $|V'| \le f(k)$, use Lemma~\ref{lemma:brute} to
  decide whether $(B',k') \in \PLang{saturation}^\ge(P)$ holds. If $|V'| >
  f(k)$, by assumption we know that $(B',k') \in
  \PLang{saturation}^\ge(P)$. In both cases, by the properties of
  reduction rules, we can output the answer for $(B',k')$ also for
  $(B,k)$. 
\end{proof}

As a final preparation, it will be useful to get two trivial cases
``out of the way'', namely graphs that have no edges or that are
complete cliques:

\begin{lemma}\label{lemma:trivials}
  For every pattern $P$ there is a $\Para\Class{AC}^0$
  algorithm that on input $(B,k)$ correctly outputs whether $(B,k) \in
  \PLang{saturation}^\ge(P)$ holds, whenever $B$ is the empty graph (has
  no edges) or is a complete clique.
\end{lemma}

\begin{proof}
  For an empty graph~$B$, any witness function can only use arcs
  from~$A^\ominus$. If there is an arc from black to black, we can
  color all vertices black and, hence, we can accept 
  the input whenever there are at least $k$ vertices. If such an arc
  does not exist, suppose there is an arc from the black color to the 
  white color in $A^\ominus$. Then, by Lemma~\ref{lemma:onlycycles},
  there must also be an arc back to the white color or a self-loop at
  the white color. In the first case color all vertices black except
  for one, in the second case color all vertices black except for two
  -- and these colorings are optimal, so we can accept the input when
  the number of vertices is $k+1$ or $k+2$, respectively. Finally, if
  there is only a self-loop at the white color in $A^\ominus$,
  everyone can and must be colored white; so we can and must accept
  iff $k=0$.

  For clique graphs $B$, the argument is exactly the same, only for
  $A^\oplus$. 
\end{proof}

With all preparations in place, our goal is now to show
$\PLang{saturation}^\ge(P) \subseteq \Para\Class{AC}^0$ for all 256
possible binary~$P$. We do not wish, of course, to go over all cases
individually -- rather, we try to deal with many of them at the same
time by making use of Lemmas \ref{lemma:symmetric}
and~\ref{lemma:onlycycles}. Indeed, by Lemma~\ref{lemma:onlycycles} we
know that $P$ cannot be ``acyclic'' and, thus, every arc must lie on some cycle in $A^\oplus \cup
A^\ominus$. We go over the different places where these cycles can
be. 

\subparagraph*{The Cases: A Self-Loop at Color Black.}
We start with the case that there is a self-loop at the color black,
that is, we consider patterns $P$ with
$(\mathrm{black},\mathrm{black}) \in A^\oplus$ or
$(\mathrm{black},\mathrm{black}) \in A^\ominus$ -- and by
Lemma~\ref{lemma:symmetric} it suffices to consider only the first
case. 

The first subcase is that there are no arcs in $A^{\ominus}$:

\begin{lemma}
  $\PLang{saturation}^\ge(P) \in \Para\Class{AC}^0$ for
  $\oplus$-superpatterns $P$ of
  \!\patterntwo{any,draw=none,overlay}{}{->}{$\oplus$}{any,draw=none,overlay}{}{any,draw=none,overlay}{}.   
\end{lemma}

\begin{proof}
  We show that $P$ is heavily saturating on large graphs (and we are
  then done by Lemma~\ref{lemma:heavy}): If $B$ contains an isolated
  vertex, we cannot assign a witness to it (since $A^\ominus =
  \emptyset$) and, thus, $B$ cannot be $P$-saturated (and, formally, a
  reduction maps the input to a trivial non-instance). Otherwise, we
  can color all vertices black and assign any adjacent vertex as
  witness. Clearly, this means that for graphs of size at least~$k$ we
  get $k$ black vertices.
\end{proof}

The second subcase is that there is an arc in $A^{\ominus}$ from black
to black:

\begin{lemma}\label{lemma:pm}
  $\PLang{saturation}^\ge(P) \in \Para\Class{AC}^0$ for superpatterns~$P$ of
  \patterntwo{any,draw=none,overlay}{}{->}{$\oplus\ominus$}{any,draw=none,overlay}{}{any,draw=none,overlay}{}.
\end{lemma}

\begin{proof}
  We can always color all vertices black in a
  $P$-saturation. Thus, $P$ is heavily saturating on graphs of size
  $k$ or more.
\end{proof}

The third subcase is that there is an arc in $A^{\ominus}$ from white
to black: 

\begin{lemma}\label{lemma:case2}
  $\PLang{saturation}^\ge(P) \in \Para\Class{AC}^0$ for
  superpatterns~$P$ of 
  \patterntwo{any,draw=none,overlay}{}{->}{$\oplus$}{any,draw=none,overlay}{}{->}{$\ominus$}.
\end{lemma}

\begin{proof}
  First suppose that there is no arc in $A^{\ominus}$ that starts at the
  black color. Then,  we \emph{must}
  color all isolated vertices white -- and we \emph{can} do so by
  coloring all non-isolated vertices black and all isolated vertices
  white (and may assume that non-isolated vertices exist, since we
  took care of a empty graph already in
  Lemma~\ref{lemma:trivials}). Thus, by removing all isolated vertices
  we get a membership equivalent graph and $P$ is thus heavily
  saturating for large graphs after this preprocessing. 

  Now suppose that there is an arc in $A^{\ominus}$ that starts at the
  black color. The case that it goes from black to black was already
  dealt with in Lemma~\ref{lemma:pm}, so it must go from black to
  white. But, then, we can color everyone black except for a single
  isolated vertex~$i$ (if it exists), making it the witness of everyone
  else and making anyone else the witness of~$i$. Thus,
  $P$ is heavily saturating on graphs with size at least $k+1$. 
\end{proof}

The fourth subcase is that there is an arc in $A^{\ominus}$ from white
to white: 

\begin{lemma}\label{lemma:case1}
  $\PLang{saturation}^\ge(P) \in \Para\Class{AC}^0$
  for superpatterns~$P$ of
  \patterntwo{->}{$\ominus$}{->}{$\oplus$}{any,draw=none,overlay}{}{any,draw=none,overlay}{}\!. 
\end{lemma}

\begin{proof}
  We may assume that there are no arcs in $A^\ominus$ that end at the
  black vertex since we already took care of these cases in the
  earlier lemmas. Thus $A^\ominus$ might only contain, in addition to
  the loop at the white color, an extra arc from black to white.

  If the extra arc is not present (so $P$ is a $\oplus$-superpattern
  of
  \patterntwo{->}{$\ominus$}{->}{$\oplus$}{any,draw=none,overlay}{}{any,draw=none,overlay}{}),
  we can and must color all isolated 
  vertices white. If there are no isolated vertices or at least two,
  we can trivially assign witnesses to them and, thus, can remove all
  isolated vertices. So we may assume that there is at most one
  isolated vertex. If there is an arc from black to white (so $P$ is a $\oplus$-superpattern
  of
  \patterntwo{->}{$\ominus$}{->}{$\oplus$}{->}{$\ominus$}{any,draw=none,overlay}{}),
  if there 
  are two or more isolated vertices, color two of them (say $i$ and
  $j$) white and make them witnesses of one another -- and color
  everyone else black, making either a black neighbor or the white~$i$
  their witness as needed.

  After the preprocessing, we get a graph with at most one isolated
  vertex. If there are no isolated vertices, we can color everyone
  black and we are done, so let $i$ be the only isolated vertex. We
  must color it white and, additionally, we must color at least one of the
  non-isolated vertices also white so that it can serve as a witness
  for~$i$. If there is at least one connected component in $B$ of
  size~$3$ or more, this is exactly what we do: Consider a spanning
  tree of the component. Color all vertices black except for $i$ and for a
  leaf $l$ of the spanning tree. Then we can assign witnesses to all the
  black vertices, while $i$ and $l$ serve as witnesses for one
  another. Finally, if all components of $B$ have size~$2$, $B$ is a
  perfect matching plus the isolated vertex~$i$. In this case, color
  everything black except for one edge and for~$i$. Then $i$ can serve
  as the witness for both vertices of the edge and either vertex can
  serve as a witness for~$i$. All told, if there are more than $k+2$
  vertices, we can color $k$ of them black. Thus, $P$ is heavily
  saturating on large graphs. 
\end{proof}

The last subcase occurs if there is an arc in $A^\ominus$ from black to
white, but there are no other arcs in $A^\ominus$. In this case, there
must be an arc in $A^\oplus$ that starts at the white vertex
(otherwise Lemma~\ref{lemma:onlycycles} would apply). This leaves two
cases, which are addressed in the following two lemmas:

\begin{lemma}
  $\PLang{saturation}^\ge(P) \in \Para\Class{AC}^0$ for
  $\oplus$-superpatterns~$P$ of 
  \patterntwo{->}{$\oplus$}{->}{$\oplus$}{->}{$\ominus$}{any,draw=none,overlay}{}\!.
\end{lemma}

\begin{proof}
  If there are no isolated vertices, everything can be colored
  black. If there are isolated vertices, we can and must color them
  black and need a witness for them. For this, we distinguish the
  following cases: First, suppose there is an isolated edge $\{x,y\}$
  (so the edge is its own connected component). In this case, color
  $x$ and $y$ white and make them witnesses of one another and let them
  be the witnesses of the isolated vertices; color everything else
  (the other non-isolated vertices) black. Second, suppose there is an
  isolated  triangle, that is, a connected component consisting of three
  vertices $x$, $y$, and~$z$. Color them white, make them witnesses of
  one another and of the isolated vertices, and color everything else
  black. Third, pick any edge $\{x,y\}$ of $B$ such that there is no
  $z$ whose neighborhood is exactly $\{x,y\}$ (we will argue in a
  moment that such an edge must exist). Color $x$ and $y$ white, make
  them witnesses of one another and of the isolated vertices. We color
  all non-isolated vertices $z$ black and can assign witnesses as
  follows: If $z$ has a neighbor other than $x$ and $y$, this
  neighbor will be black and can serve as a witness
  for~$z$. Otherwise, either $x$ or $y$ is \emph{not} a neighbor of
  $z$ and, thus, can serve as a white witness of the black $z$ via the
  $\ominus$-arc from black to white. All told, we see that $P$ is
  heavily saturating.

  It remains to argue that we always find an edge $\{x,y\}$ such that
  no $z$ has only these two vertices as its neighborhood. However,
  suppose this were not the case, so for every edge $\{x,y\}$ there is
  a $z$ having exactly $\{x,y\}$ as its neighbors. Then $z$ has
  degree~$2$. Consider the edge $\{x,z\}$. By assumption, there is
  once more a vertex $z'$ that has exactly $x$ and $z$ as its
  neighbors. However, the only two neighbors of $z$ were $x$ and $y$,
  so $z'=y$ must hold. Thus, the only neighbors of $y$ are $x$ and
  $z$. By considering the edge $\{y,z\}$ we also get that the only
  neighbors of $x$ are $y$ and $z$. All told, $x$, $y$, and $z$ must
  form an isolated triangle -- which we ruled out earlier.
\end{proof}

\begin{lemma}\label{lemma:case3}
  $\PLang{saturation}^\ge(P) \in \Para\Class{AC}^0$ for
  $P ={}$%
  \patterntwo{any,draw=none,overlay}{}{->}{$\oplus$}{->}{$\ominus$}{->}{$\oplus$}.
\end{lemma}

\begin{proof}
  We can reject any $B$ that consists only of isolated
  vertices and we also reject $B$ if it consists only of isolated
  vertices plus a single edge $\{u,v\}$: The isolated vertices would
  need to be colored black and at least one of $u$ and $v$ would need
  to be colored white, say~$u$. But, then, $v$ would have to be
  colored black as it is the 
  only possible witness for~$u$. This leaves us without any possible
  witness for $v$ since, as a black vertex, it would need to be
  connected by an edge to a black vertex (but $u$ is white) or by a
  non-edge to a white vertex (but everyone else is black).

  In all other cases, first consider the situation that there are
  no isolated vertices. Then we can color everyone black. Otherwise, we
  must color all isolated vertices black and pick some non-isolated
  $v$ to serve as their white witness. As in the proof of
  Lemma~\ref{lemma:case1}, if there is a connected component of size
  at least~$3$ in~$B$, pick a leaf of a spanning tree of this
  component as~$v$. Color $v$ white and everyone else black. The
  witness of $v$ is then any neighbor of $v$ in~$B$. Finally, if all
  connected components are single edges, $B$ is a matching plus some
  isolated vertices. We already ruled out the case of a single edge,
  so let $\{u,v\}$ and $\{x,y\}$ be two edges of the matching. Color
  $u$ and $x$ white, everyone else black. Then the white $u$ can serve
  as a witness of all isolated vertices as well as of the black~$y$ to
  which it is not connected. Next, the white $x$ can
  serve as the witness of the black~$v$. The witness of the white $u$
  is the black~$v$ to which it is connected, and the  witness of the
  white $x$ is the black~$y$. All told, $P$ is heavily saturating
  after the preprocessing. 
\end{proof}

\subparagraph*{The Cases: A Self-Loop at Color White.}

We now consider the case that there is a self-loop at color white and,
as before, it suffices that consider the case that there is a
$\oplus$-arc. For the following cases, we no longer need to consider
any situations where there is a self-loop at color black since we
already took care of them earlier.

The remaining cases start with one a self-loop at the white color and
no other arcs:

\begin{lemma}
  $\PLang{saturation}^\ge(P) \in \Para\Class{AC}^0$ for
  $P \in\{\text{\patterntwo{->}{$\oplus$}{any,draw=none,overlay}{}{any,draw=none,overlay}{}{any,draw=none,overlay}{}},
  \text{\patterntwo{->}{$\oplus\ominus$}{any,draw=none,overlay}{}{any,draw=none,overlay}{}{any,draw=none,overlay}{}}
  \}$.
\end{lemma}

\begin{proof}
  Everything must clearly be colored white, so there can only be a
  solution for $k=0$, but then Lemma~\ref{lemma:slice} applies. 
\end{proof}

The case that there are two arcs from black to white and a self-loop
at the white color is also trivial:

\begin{lemma}\label{lemma:both}
  $\PLang{saturation}^\ge(P) \in \Para\Class{AC}^0$ for superpatterns
  $P$ of 
  \patterntwo{->}{$\oplus$}{any,draw=none,overlay}{}{->}{$\oplus\ominus$}{any,draw=none,overlay}{}\!.
\end{lemma}

\begin{proof}
  For a basic input graph~$B$ let $\{u,v\} \in E$ (we can assume
  that such an edge exists by Lemma~\ref{lemma:trivials}). Color 
  $u$ and $v$ white, make them witnesses of one another, and color
  everyone else black, making $u$ (or $v$) their witness. Clearly, $P$
  is heavily saturating for graphs with at least $k+2$ vertices.
\end{proof}

If there is an arc from white to black, there must also be an arc
backward from black to white since, by Lemma~\ref{lemma:onlycycles} we
cannot ``stop'' at the black color and since we already ruled out the
cases having a self-loop at the black color. Thus, it suffices to
consider only the two cases from the next two lemmas:

\begin{lemma}
  $\PLang{saturation}^\ge(P) \in \Para\Class{AC}^0$ for superpatterns
  $P$ of 
  \patterntwo{->}{$\oplus$}{any,draw=none,overlay}{}{->}{$\oplus$}{any,draw=none,overlay}{}\!.
\end{lemma}

\begin{proof}
  We may assume that in~$P$ there is no arc in $A^\ominus$ from black
  to white (otherwise Lemma~\ref{lemma:both} applies) nor one from
  black to black (self-loops at color black have already been taken
  care or).

  For an input $B$, we start by taking care of the isolated
  vertices. Suppose there is a 
  non-empty set $I$ of isolated vertices (but $I \neq V$ by
  Lemma~\ref{lemma:trivials}). How we can proceed, depends on 
  which arcs are present in~$A^\ominus$. If there are no arcs in
  $A^\ominus$ at all, we can and must reject the graph (there are no
  witnesses for any members of~$I$). If there is an arc, it must start
  at the color white. This allows us to simply remove all vertices
  in~$I$ from the graph as a reduction rule: They \emph{must} be
  colored white, but any weight-$k$ coloring of the remaining
  graph can be extended to a coloring of the whole graph: If there is
  an arc in $A^\ominus$ from white to white, assign any white vertex
  in the coloring of the reduced graph as a witness (which must exist
  as all arcs start or end at color white); and if there is an arc
  from white to black, assign any black vertex as a witness. For $k>0$
  such a vertex must always exist; and for $k=0$ we can apply
  Lemma~\ref{lemma:slice}. 

  We now have a graph $B$ that contains no isolated vertices. Consider
  the set $M$ of all isolated edges (so edges where both endpoints
  have degree~$1$). Once more, we can apply different kinds of
  reductions: If there are no arcs in $A^\ominus$, we must color both
  endpoints of any $\{u,v\} \in M$ white since they are the only
  possible witnesses for one another and if we color one of them black,
  we have no witness for the other one. However, we also \emph{can}
  color all of them white as they can serve as witnesses for one
  another. Thus, we can remove all of $M$ and all vertices in it from
  $B$ as a reduction step. Next, if there is an edge from white to
  white in $A^\ominus$, for each $\{u,v\} \in M$ we can color one of
  them (say, $u$) black and the other white. Then the white $v$ is a
  witness for the black~$u$ and any white vertex in the remaining
  graph (which must exist if it can be $P$-saturated at all) can serve
  as the witness for~$v$ (as $v$ is not connected to that vertex, it
  is only connected to~$u$). Likewise, if there is an arc from white
  to black, we can use any black vertex other than $u$ as the witness for~$v$. In either
  case, however, we cannot color \emph{both} $u$ and $v$ black since
  this would rob us of witnesses for them. All told, if $|M| \ge k+1$ we
  can accept (just color one vertex from the first $k$ edges in $M$
  black and everyone else white), and otherwise we can remove $M$~from
  the graph and lower $k$ by $|M$|.

  We now have a graph without isolated vertices or edges, that is,
  every connected component has size at least~$3$. We claim that,
  then, if $B$ is sufficiently large, it can be $P$-saturated with
  weight~$k$. To show this, let $C_1$, \dots, $C_l$ be the connected
  components of~$B$ and let $T_1$, \dots, $T_l$ be spanning trees of
  the~$C_i$. In each $T_i$, call some leaf node $r_i$ the \emph{root}
  of the tree. For a node $v$ in $T_i$ other than $r_i$ let $w(v)$ be
  the next vertex on the path from $v$ to the root. For the root
  $r_i$, let $v_i$ be its neighbor and define $w(r_i) = v_i$. Note that
  $v_i$ is not a leaf as all components have size $3$ or more. Also note
  that each $T_i$ has at least one leaf other than its root. We call
  leaves that are not roots \emph{proper leaves}.

  First, suppose that there are at least $k$ trees $T_i$. Then we can
  color all proper leaves in the $T_i$ black (this colors at least $k$
  vertices black) and color everything else white. The witness function~$w$
  that we already defined is now correct: Leaves have a white vertex as
  their witness since the parent of a leaf is not a leaf itself. The
  roots, which are special, are also connected by an edge to a white
  vertex $v_i$ since the $v_i$ are not leaves.

  Second, suppose that there is a tree with at least $k$ proper
  leaves. By the same argument as before, coloring all proper leaves
  black and everyone else white yields a valid $P$-saturation of
  weight at least~$k$.

  Third, suppose that there is a tree with less than $k$ proper
  leaves, but more than $3k^2+2k$ vertices. Then this tree must contain a
  path $(u_1,\dots,u_{3k})$ of length at least $3k$ such that all
  vertices on the path have degree $2$~(with respect to the tree) and
  that does not include the parent $v_i$ of the root. The
  reason is that any vertex of degree larger than~$2$ in a tree
  implies the existence of another leaf, so we get a bound of $k$ on
  the number of vertices of degree $1$ and also on degree larger
  than~$1$. Furthermore, the degree-2 vertices can form at most $k$
  paths -- so if there are $3k^2$ many of them, a path of length $3k$
  must be present.

  Let the witnesses along the path point to the smaller index, so
  $w(u_2) = u_1$, $w(u_3) = u_2$ and so on. Color the vertices as
  follows: Color $u_{3i-2}$ for $i \in \{1,\dots,k\}$ black, color all
  other vertices white. Modify the witness function that we described
  earlier as follows: Set $w(u_{3i-2}) \coloneq u_{3i-1}$ and
  $w(u_{3i-1}) \coloneq u_{3i}$ for $i \in \{1,\dots,k\}$, but leave
  $w(u_{3i})$ unchanged at $u_{3i-1}$. To see that this yields
  a correct witness function, not that the black vertices $u_{3i-2}$
  have the white witnesses $u_{3i-1}$, the white vertices $u_{3i-1}$
  have the white witnesses $u_{3i}$, and the white vertices $u_{3i}$
  have the white witnesses $u_{3i-1}$. Furthermore, vertices not on
  this path are white and will only use (at best) $u_{3k}$ as their
  witness, which is white.

  All told, we can always $P$-saturate $B$ with $k$ black vertices
  when $B$ has more than $3k^3+2k^2$ vertices (it either consists of
  at least $k$ connected components or, if not, one of these
  components has $k$ leaves or, if not, it must have size at least
  $(3k^3+2k^2)/k = 3k^2+2k$). This means that $P$ is heavily
  saturating for graphs of size $3k^3+2k^2$ after preprocessing.
\end{proof}

\begin{lemma}
  $\PLang{saturation}^\ge(P) \in \Para\Class{AC}^0$ for superpatterns
  $P$ of 
  \patterntwo{->}{$\oplus$}{any,draw=none,overlay}{}{->}{$\ominus$}{any,draw=none,overlay}{}\!.
\end{lemma}

\begin{proof}
  For this proof, it will be convenient to apply
  Lemma~\ref{lemma:symmetric} once, so assume that $P$ is actually 
  a superpattern of
  \patterntwo{->}{$\ominus$}{any,draw=none,overlay}{}{->}{$\oplus$}{any,draw=none,overlay}{}\!. 
  Our objective is to show that $P$ is still heavily saturating after
  preprocessing. 
    
  Again by Lemma~\ref{lemma:trivials}, we can assume that there is at
  least one non-edge in~$B$, that is, that at least one vertex is not
  \emph{universal} (connected to all other vertices).  
  We start by taking care of the set~$U$ of universal vertices
  in~$B$. If $|U| \ge k$, (this threshold check 
  be done in $\Para\Class{AC}^0$), we can accept~$B$ since it can be
  $P$-saturated as follows: Color the (at 
  least $k$ many) universal vertices black and all other vertices 
  white. Since there is at least one non-edge
  $\{u,v\} \notin E$, we can assign the following witnesses: All universal
  (and, thus, black) vertices get~$u$ as their witness (which is
  white and connected to every universal vertex, so the witness
  property is fulfilled for the universal vertices). The non-universal
  vertices get any of their non-neighbors as witness (which are all
  white and the witness property is once more fulfilled). Next, if $0 < |U|
  < k$, apply the following reduction rule: Remove $U$ and all
  adjacent edge from the graph and set the new parameter $k' = k -
  |U|$. The correctness follows since if we can $P$-saturate the 
  reduced graph with some witness function and some coloring that has
  at least $k'= k - |U|$ black vertices, we can extend this coloring
  to a coloring for the full graph~$B$ by coloring all of the universal
  vertices that we removed black and by assigning to them as a witness
  any white vertex of the reduced graph (which must exist). This gives
  us a $P$-saturation of the original~$B$ by a coloring of weight at
  least $k'+|U| \ge k$. In the following we may now assume that there
  are no universal vertices.
  
  Suppose there is a vertex $v$ of degree at least~$k$. Then we get a
  weight-$k$ coloring by coloring the neighbors of~$v$ black and
  everyone else (including~$v$) white. Then the white~$v$ is a legal
  witness for all of its (black) neighbors and also for all its
  (white) non-neighbors. A witness for $v$ is any (white) non-neighbor
  of~$v$ (which must exist as $v$~is not universal).

  At this point, if there are any isolated vertices, we remove
  them (this is a reduction rule). The rule is correct since isolated
  vertices \emph{must} be colored white (there is no arc starting at
  color black in $A^\ominus$, otherwise we would already have handled
  the pattern in Lemma~\ref{lemma:both}) and they \emph{can} be colored white as
  long as there is another white vertex in the resulting graph (which
  there must be, if a $P$-saturation is possible at all). 
  
  Now suppose that $B = (V,E)$ has at least $2k^2$ edges in~$E$. Then
  there must exist a matching $M \subseteq E$ of size at least~$k$
  since we can greedily construct such a matching (pick an edge
  $\{u,v\}$ in~$E$, add it to $M$, and remove the at most $2k$ edges
  that contain $u$ or~$v$). Using the resulting $M =
  \{\{u_1,v_1\},\dots,\{u_k,v_k\}\}$ we can then $P$-saturate $B$ with
  weight~$k$ by coloring exactly the $u_i$ black and all other
  vertices white. We can assign the following witnesses: Each black
  $u_i$ gets the white $v_i$ as its witness. Any white vertex $x$ is
  \emph{not} connected to at least one vertex $v \in
  \{v_1,\dots,v_k\}$ (because of the degree bound). Hence, we can set
  $w(x) = v$.

  To conclude, we find that whenever $B$ has at least $4k^2$ vertices,
  it will have at least $2k^2$ edges (since there are no isolated
  vertices) and hence a $P$-saturating weight-$k$ coloring. We get
  that $P$ is heavily saturating for graphs of size $4k^2$ or larger
  (after preprocessing).
\end{proof}

\subparagraph*{The Cases: No Self-Loops.}

The last remaining cases are patterns in which there are no
self-loops at either color. This means that there is an arc from black to
white in $A^{\oplus} \cup A^{\ominus}$ and an arc from white to black
also in this set. The first of two lemmas will show that if we find
two such arcs in $A^\oplus$, we are done; and by
Lemma~\ref{lemma:symmetric} this implies the same for the case that
there are two arcs in $A^\ominus$. The remaining two cases are that
one of the arcs comes from $A^\oplus$ and the other from $A^\ominus$
and there are no other arcs. Again by symmetry, it suffices to
consider the case of an $A^\oplus$ arc from white to black and an
$A^\ominus$ arc from black to white. This case, by far the most
complicated one, is addressed in the final lemma of this section.

\begin{lemma}\label{lemma:pp}
  $\PLang{saturation}^\ge(P) \in \Para\Class{AC}^0$ for superpatterns
  $P$ of \patterntwomixed{->}{$\oplus$}{->}{$\oplus$}.
\end{lemma}

\begin{proof}
  Let us first deal with the existence of an isolated vertex~$i$
  in~$B$. If there are no arcs in $A^{\ominus}$, we can directly
  reject the input as we cannot assign a witness to~$i$. However,
  then, there needs to be at least
  one $\ominus$-arc and, since we already ruled out self-loops, must
  go from black to white or from white to black. Furthermore, since
  \emph{only} $\ominus$-arcs can serve as witnesses, we must actually
  have both $\ominus$-arcs from white to black and from black to white
  (or we can immediately reject). But, then, we can color all vertices 
  black except for $i$ and make the white, isolated~$i$ the witness of
  all black vertices and make any other (then automatically black)
  vertex the witness for~$i$.

  It remains to deal with graphs without isolated vertices. First
  suppose that $B$ has least $k$ connected components of size at
  least~$2$. Then for each connected component~$C$ of~$B$, choose any
  vertex $c\in C$ and color it black. For each vertex $v$ of the
  component, 
  color it white if it has odd distance from $c$, and black otherwise. Setup the
  witness function $w$ as follows: Map $c$ to any of its neighbors
  (which must exists as there are no isolated vertices). Map each
  vertex $v$ in the component to one of its neighbors that has distance~$1$ less
  from $c$. Clearly, such a neighbor must exist and it will have the opposite
  color from~$v$. With this strategy, we get a valid coloring of
  weight at least~$k$. Second, suppose one connected component has at
  least $2k$ vertices. Then the same strategy leads to a valid
  coloring and either this coloring or the inverted coloring (black
  vertices become white and \emph{vice versa}) will be a coloring of
  weight at least~$k$. Third, if there are less than $2k$ components
  in~$B$, each of size less than $2k$, the size of $B$ is bounded in
  terms of the parameter. All told, $P$ is heavily saturating if $B$
  has size at least $4k^2$.
\end{proof}

\begin{lemma}\label{lemma:self}
  $\PLang{saturation}^\ge(P) \in \Para\Class{AC}^0$
  for $P={}$\patterntwomixed{->}{$\ominus$}{->}{$\oplus$}.
\end{lemma}

\begin{proof}
  As in previous cases, we will present a series of reduction rules to
  tackle the problem. The main objective will be to arrive at a
  restricted class of basic graphs, which we will call \emph{peeled}
  graphs (the reason for the name will become clear later on). To
  define them, recall 
  that an \emph{isolated vertex} in a graph is a vertex without
  neighbors, and, symmetrically, a \emph{universal vertex} is a vertex
  such that all (other) vertices are neighbors. A \emph{peeled} graph
  is a graph without isolated and without universal vertices. Our
  interest in the graphs lies in the following claim:
  \begin{claim}\label{claim:iso-uni}
    If a peeled $B$ can be $P$-saturated (at all), then $B$ can be
    $P$-saturated via a coloring having weight at least $\lfloor
    n/2\rfloor$. 
  \end{claim}
  \begin{proof}
    Let $w \colon V \to V$ be a witness function and 
    $c \colon V \to C$ be a coloring that $P$-saturates $B$ such that $c$
    has maximal weight. If $\operatorname{weight}(c) \ge \lfloor n/2\rfloor$ we
    are done, so assume that the number of vertices colored black is
    strictly less than $\lfloor n/2\rfloor$. Let $W$ denote the set of white
    vertices (so $W = \{v \in V \mid c(v) = \mathrm{white}\}$) and let
    $K$ denote the set of black vertices.

    We claim that for each white vertices $a \in W$ there is some black 
    vertex $b \in K$ with $\{a,b\} \notin E$: Suppose this were not
    the case for some $a \in W$. Then $a$ is connected to all $b \in
    K$. Since $a$ is not a universal vertex, $a$ cannot also be
    connected to all other white vertices, so there is another white
    vertex $a'\in W$  with $\{a,a'\} \notin E$. We can now change the
    color of $a$ from white to black, if we also change the witness of
    $a$ to be $a'$: As a black vertex, we need its witness $a'$ not be
    connected to $a$, which is exactly the case. Since $a$ was
    connected to all black vertices, it is also impossible that $a$
    was a witness for any of these black vertices. All told, we have
    constructed a saturating coloring of weight \emph{one larger than
    before,} contradicting the assumption that the weight was maximal.

    Our next claim is that for each black vertex $b \in K$ there is
    some white vertex $a \in W$ with $\{a,b\} \in E$: Suppose once
    more that this were not the case for some $b \in K$. Since $b$ is
    not isolated, we know that there is a $b' \in K$ with $\{b,b'\}
    \in E$. Consider the set $X = \{w(b'') \mid b'' \in K, b'' \neq
    b\}$ of (white) witnesses of the black vertices other
    than~$b$. Since $|K|  < \lfloor n/2\rfloor$, we see that $|X| <
    \lfloor n/2\rfloor - 1$. In particular, $|W \setminus X| \ge \lfloor
    n/2\rfloor -(\lfloor n/2\rfloor - 1) + 1 = 2$, that is, there are
    at least two white vertices that are not witnesses of the black
    vertices (other than~$b$). Consider the following recoloring and
    rewitnessing: Color $b$ white, but color all vertices in $W
    \setminus X$ black. Let $b$ be the (now white) witness of the (now
    black) vertices in $W \setminus X$ (there are no edges between $b$
    and any vertex in $W$ by assumption), let $b'$ be the (still
    black) witness of (the newly white)~$b$, and observe that
    recoloring the vertices in $W \setminus X$ does not rob us of any  
    of the vertices used as witnesses for the vertices in~$K$. Since
    we made one vertex white that used to be black, but made at least
    two white vertices black, we now have a new saturating coloring of
    a larger weight than before. This contradicts our assumption.

    Our third claim is that we can now invert the coloring and still
    obtain a valid $P$-saturation of~$B$: The reason is simply that
    by the first claim, for every (now black) vertex $a \in W$ there is a
    (now white) vertex $b \in B$ such that $\{a,b\} \notin E$ and this
    $b$ is a permissible witness for $a$; and by the second claim, for
    every (now white) vertex $b \in K$ there is a (now black) vertex
    $a \in W$ with $\{a,b\} \in E$, so $b$ is a permissible witness
    for~$a$. However, the inverted coloring has larger weight than the
    original coloring, leading once more to a contradiction -- so, all
    told, the only case that remains is that the saturating coloring
    of maximal weight has weight at least $\lfloor n/2\rfloor$.
  \end{proof}
  Clearly, if we can come up with a reduction rule to ``peel'' any
  graph~$B$, the claim tells us that $P$ is heavily saturating for
  graphs of size at least $2k$. Let us make this observation explicit
  for future reference:
  \begin{claim}\label{claim:peeled}
    There is a $\Para\Class{AC}^0$ algorithm that for any $(B,k)$ where
    $B$ is peeled, correctly decides
    membership in $\PLang{saturation}^\ge(P)$. 
  \end{claim}    
  To handle the case that $B$ has
  isolated vertices or universal vertices (it clearly cannot have both
  at the same time), at first sight it seems easy enough to delete
  them (here, $B - \{v\}$ 
  is obtained from $B$ by simply deleting $v$ and all adjacent edges):  
  \begin{claim}\label{claim:universal}
    Let $v$ be a universal vertex of~$B$. Then $(B,k) \in
    \PLang{saturation}^\ge(P)$ iff $(B - \{v\},k) \in
    \PLang{saturation}^\ge(P)$.
  \end{claim}
  \begin{proof}
    Observe
    that any $P$-saturation \emph{must} color $v$ white since it will
    always be connected to its witness by an edge. Conversely, any
    $P$-saturation of $B - \{v\}$ can be extended to one of $B$ by
    coloring $v$ white and making any black vertex of $B$ its witness
    (and such a vertex must exist).
  \end{proof}
  \begin{claim}\label{claim:existential}
    Let $v$ be an isolated vertex of~$B$. Then $(B,k) \in
    \PLang{saturation}^\ge(P)$ iff $(B - \{v\},k-1) \in
    \PLang{saturation}^\ge(P)$.
  \end{claim}
  \begin{proof}
    For this claim, just note that $v$ \emph{must} be colored black
    since there is no edge to its witness -- and any coloring of $B
    - \{v\}$ \emph{can} be extended to one of $B$ by making any white
    vertex in $B - \{v\}$ the witness of the black $v$.
  \end{proof}

  The two claims suggest simple reduction rules: Simply remove all
  universal vertices in parallel (without changing $k$) and then remove all
  isolated vertices (while lowering $k$ by the number of removed
  vertices). However, this approach does not quite work: After
  removing the isolated vertices, there might be new
  universal vertices (that were not universal in the original
  graph). And if we then remove these vertices in parallel, there
  might now be new isolated vertices, and so on. Each time we
  remove isolated vertices, we lower $k$ by at least $1$ and, thus, we
  will be done after at most $k$ rounds -- but a straightforward
  implementation will yield a $\Para\Class{AC}^{0\uparrow}$ algorithm
  rather than the desired $\Para\Class{AC}^{0}$
  algorithm. Furthermore, it is not hard to construct graphs in which
  we do, indeed, need a non-constant number of rounds of removals before
  we can decide whether the graph really can be $P$-saturated with
  enough black vertices. To overcome these difficulties, we need a way
  of identifying the vertices that will be removed in a large number
  of vertex removal rounds directly. Fortunately, this is possible,
  but needs some machinery.

  The process of alternating between removing universal and isolated
  vertices until a peeled graph remains can be thought of as, well, a
  \emph{peeling} of the graph: We ``peel away'' layers of isolated
  vertices and universal vertices until a ``peeled graph'' remains. To
  get a better handle on this process, some notations will be useful:
  For a basic graph $B = (V,E)$ let $S = (S_1,S_2,S_3,\dots,S_l)$ be a 
  sequence of non-empty, pairwise disjoint subsets of~$V$. Clearly, such a
  sequence is necessarily finite. Let us write
  $\operatorname{peel}_{\cdots S_m}(B)$ for~$B$ with all vertices (and adjacent
  edges) removed in $\bigcup_{j=0}^m S_j$ and
  $\operatorname{peel}_{S}(B)$ for $\operatorname{peel}_{\cdots S_l}(B)$.
  
  \begin{definition}
    A \emph{peeling of~$B$} is a sequence $S = (I_0,U_1,\dots)$ such
    that 
    \begin{enumerate}
    \item $I_j$ is the non-empty set of isolated vertices
      of $\operatorname{peel}_{\cdots U_j}(B)$, and
    \item $U_j$ is the non-empty set of universal vertices
      of $\operatorname{peel}_{\cdots I_{j-1}}(B)$.
    \end{enumerate}
    A peeling is \emph{maximal} if there is no longer peeling of~$B$. 
  \end{definition}

  The definition is clearly tailored to the case that the peeling
  process starts with isolated vertices (that is, that $B$ contains no
  universal vertices). Fortunately, we can easily ensure that this is
  always the case: If necessary, we reduce $B$ by invoking
  Claim~\ref{claim:universal} \emph{once} to remove all universal
  vertices. For such graphs, the definition now corresponds to the
  process of alternatively peeling away the isolated and universal
  vertices from~$B$, resulting in a well-defined maximal
  peeling~$S$. Crucially, $\operatorname{peel}_S(B)$ is then a peeled
  graph.

  As mentioned earlier, working directly with peelings is difficult
  when we wish to construct constant depth circuits: The natural way
  of computing $\operatorname{peel}_{\cdots U_j}(B)$ leads to a circuit
  of depth $O(k)$ and, hence, more than we are allowed. Even worse,
  the length of the sequence might be unbounded in terms of~$k$,
  meaning that the straightforward computation of the peeled graph
  $\operatorname{peel}_S(B)$ will use a depth that is not bounded in
  terms of~$k$ (let alone be constant).

  To overcome these difficulties, we use the notion of \emph{twins:}   
  We call two vertices $u$ and $v$ of~$B$ \emph{twins}, if their
  neighborhoods are identical (except for $u$ and $v$ themselves), so
  $\{ w \mid \{u,w\} \in E, w \notin \{u,v\}\} = \{ w \mid \{v,w\} \in
  E, w \notin \{u,v\}\}$. Let us write $u \equiv v$ when $u$ and $v$
  are twins and note that, indeed, this an equivalence relation
  on~$V$. In particular, each equivalence class of $\equiv$ is either
  a clique of $B$ or an independent set of~$B$.

  Let $B'= (V',E')$ be the graph that results from $B$ where we replace each
  equivalence class by a single vertex. More precisely (in order to be
  able to implement this operation in $\Class{AC}^0$), we simply
  remove from $B$ all vertices (along with their adjacent edges) that
  have a twin that comes earlier in the input, keeping only one vertex
  per equivalence class. For a vertex $v \in V$ let $v_{\mathrm{rep}}$
  denote the representative that is kept, that is, the twin of~$v$
  that comes earliest in the input.
  The advantage of working on $B'$ rather than $B$ lies in the
  following observation: 
  \begin{claim}\label{claim:peelingtotwin}
    Let $(I_0,U_1,I_1,U_2,\dots)$ be the maximal peeling of~$B$ (as
    always, we assume $B$ has no universal vertices). Then
    $(\{i_0\}, \{u_1\}, \{i_1\}, \{u_2\}, \dots)$ is the maximal
    peeling sequence of $B'$, where each $i_j$ is the representative
    of any (and all) vertices in $I_j$ and $u_j$ is the representative
    of any (and all) vertices in $U_j$.
  \end{claim}
  \begin{proof}
    First consider the case that $B$ is already peeled, there are no
    isolated vertices in~$B$. Then, there are also no isolated
    vertices in~$B'$ (because such a vertex would also be isolated
    in~$B$).

    Now suppose that $B$ is not yet peeled, but has no
    universal vertices. Then $I_0$ is the non-empty set of isolated
    vertices of~$B$. All vertices in $I_0$ are twins and, thus, in
    $B'$ they are all represented by their
    representative~$i_0$. Furthermore, $i_0$ is isolated in~$B'$ and
    there are no other isolated vertices in $B'$ (since they would be
    twins of $i_0$ in the original~$B$).

    Let $U_1$ be the set of universal sequences of $B - I_0$. If this
    set is empty (meaning that the peeling of $B$ ends at $I_0$),
    there is also no universal vertex $u$ in $B' - \{i_0\}$: Such a
    vertex would be the representative of a clique~$U$ in $B - I_0$
    and hence be connected to all other vertices of $B - I_0$, which
    we just ruled out. Thus, the peeling of $B'$ stops 
    synchronously at $\{i_0\}$. On the other hand, if $U_1$ is not
    empty, its members form a clique and they are all twins. Then
    their representative $u_1$ is a universal vertex in $B'-\{i_0\}$
    since it is clearly connected to all vertices in that
    set. Furthermore, there are no other universal vertices since they
    would have been twins of~$u_1$. This means that the peeling of
    $B'$ continues with $\{u_1\}$.

    For the rest of the sequence, we can repeat the arguments from the
    above two paragraphs in an alternating manner.
  \end{proof}

  The above claim allows us to switch to peelings in which in each
  step exactly one vertex is peeled away. However, it is still not
  obvious how we can compute it using an $\Class{AC}^0$ circuit. The
  trick is the following observation:
  \begin{claim}\label{claim:degrees}
    Let $(\{i_0\}, \{u_1\}, \{i_1\}, \{u_2\}, \dots)$ be the maximal
    peeling of $B' = (V',E')$. Then each vertex $i_j$ has degree exactly $j$
    in~$B'$ and is connected exactly to the vertices $u_1$,
    \dots,~$u_j$. Furthermore, all vertices in $V' \setminus
    \{i_1,\dots,i_j\}$ have degree at least $j$ in~$B'$. 
  \end{claim}
  \begin{proof}
    The neighborhood of  $i_j$ is, indeed, $\{u_1,\dots,u_j\}$ by
    construction, which trivially implies the degree claim. To see
    that the other vertices have degree at least~$j$, just observe
    that they are all connected to $u_1$ to~$u_j$.  
  \end{proof}
  We get an interesting corollary of this claim:
  \begin{claim}\label{claim:structure}
    Suppose that the peeling sequence of~$B'$ has length at least
    $2l+1$. Then:
    \begin{enumerate}
    \item For each $j \in \{0,\dots,l\}$ there is exactly one vertex
      of degree $j$ in~$B'$. Let us call it~$i_j$.
    \item For each $j \in \{1,\dots,l\}$ there is exactly one vertex
      connected to $i_j$ but not to $i_{j-1}$. Let us call it~$u_j$.
    \item For $j \in \{0,\dots,l\}$, the vertex~$i_j$ is connected
      exactly to $u_1$, \dots, $u_j$.
    \item For $j \in \{1,\dots,l\}$, the vertex~$u_j$ is connected 
      exactly to the vertices in $V'\setminus\{i_0,\dots,i_{j-1}\}$.
    \item The sequence $(\{i_0\}, \{u_1\}, \{i_1\}, \{u_2\}, \dots,
      \{i_l\})$ is a prefix of the maximal a peeling of $B' = (V',E')$
      (and, hence, a peeling itself).
    \end{enumerate}
  \end{claim}
  \begin{proof}
    All of the properties follow directly from the previous
    claim concerning the degrees of the vertices. 
  \end{proof}

  The important aspect of the above claim is, of course, that the
  first four properties can easily be tested using parameter-dependent
  thresholds, meaning that we can do all the tests using
  $\Para\Class{AC}^0$ circuits. We can now solve the original problem
  as follows:

  On input $(B,k)$ where $B$ has no universal vertices (recall that we
  can ensure this), first check whether $B$ also has no isolated
  vertices (and is, hence, peeled). If so, use 
  Claim~\ref{claim:peeled} to decide whether $(B,k) \in
  \PLang{saturation}^\ge(P)$ holds. Otherwise, do the following in
  parallel for all $l \in \{0,\dots,k\}$: Check whether for the given
  $l$ the first four items of Claim~\ref{claim:structure} hold. Let
  $m$ be the largest~$l$ for which this is the case.

  By the last item of Claim~\ref{claim:structure} we know that the
  peeling of $B'$ starts with  $(\{i_0\}, \{u_1\},\penalty0 \{i_1\},\penalty0 \{u_2\},
  \dots,\penalty50 \{i_m\})$. Note that, crucially, a $\Para\Class{AC}^0$
  circuit can compute each element $i_j$ and $u_j$ of this sequence
  since Claim~\ref{claim:structure} then only uses  
  parameter-dependent thresholds to identify the vertices. Let $(I_0,
  U_1, I_1, U_2, \dots, I_m)$ be the start of the peeling of~$B$ and
  recall from Claim~\ref{claim:peelingtotwin} that each $i_j$ is the
  representative of all vertices in $I_j$ and similarly for the $u_j$
  and~$U_j$. 
  
  Consider the graph $C = \operatorname{peel}_{\cdots I_m}(B)$. We can
  compute this graph in $\Para\Class{AC}^0$ since it is simply the
  graph induced in $B$ on the set of all vertices that are not twins
  of any vertex in $\{u_1,\dots,u_m,i_0,\dots,i_m\}$. Observe that
  $C$ has the following property:
  \begin{align}
    &\text{\itshape $C$ can be $P$-saturated by a
      coloring of weight~$w$} \tag{$\ast$}\label{property:ast}\\
    \text{\itshape iff\ \ } & \text{\itshape $B$ can be $P$-saturated by a coloring of
    weight $\textstyle w + \sum_{j=0}^m |I_j|$.}\notag
  \end{align}
  This is because of Claims \ref{claim:universal}
  and~\ref{claim:existential} by which 
  peeling away universal vertices does not change the weight of the
  possible colorings that $P$-saturate $B$ while peeling away $|I_j|$
  isolated vertices results in a graph whose $P$-saturating colorings
  have exactly $|I_j|$ black vertices less than those of the unpeeled
  graph.

  The final two arguments are the following: We check whether
  $\sum_{j=0}^m |I_j| \ge k$ holds, which is a threshold check that
  can be done in $\Para\Class{AC}^0$. Note that this test will always
  be true if $m =k$ since $|I_j| \ge 1$ holds as we always peel away at
  least one isolated vertex. Now, in this case, by Property \eqref{property:ast},
  $C$ can be $P$-saturated at all iff $B$ can 
  be $P$-saturated with some weight at least $k$. As we can check $C  
  \in \Lang{saturation}(P)$ in $\Class{FO} = \Class{AC}^0$, we 
  are done in this case.

  Now suppose  $\sum_{j=0}^m |I_j| < k$ and, hence, $m < k$. Then 
  $m+1$ did \emph{not} pass the tests of Claim~\ref{claim:structure}
  and, hence, either $(\{i_0\}, \{u_1\}, \{i_1\}, \{u_2\}, \dots,
  \{i_m\})$ is the maximal peeling of $B'$ or this sequence with one
  more universal peeling added at the end. In particular, $C$ or $C$
  with one more universal peeling is a \emph{peeled graph} (we cannot
  extend the peeling sequence). By Property \eqref{property:ast} we get that the
  peeled basic graph $C$ can be $P$-saturated with a coloring of
  weight $k'\coloneq k - \sum_{j=0}^m |I_j|$ iff $B$ can be $P$-saturated
  with a coloring of weight~$k$. By Claim~\ref{claim:peeled}, we get
  the claim.
\end{proof}

Putting it all together, we conclude:\label{page:lemma-ae-last}

\begin{proof}[Proof of Lemma~\ref{lemma:basic-ae}]
  In the course of this section, we have shown that for all binary
  patterns~$P$ we have 
  $\PLang{saturation}^\ge(P) \subseteq
  \Para\Class{AC}^0$. By Lemma~\ref{lemma:fd-to-satu}, we then also
  have 
  $\PClass{FD}_{\mathrm{basic}}(E^\ge ae) \subseteq
  \Para\Class{AC}^0$. 
\end{proof}

\end{scope~}

%
%
%
%
\subsection[The Case Greater-Than Case for aa and Basic Graphs]
{The Case \textit{E$_{\text{\upshape 1}}^{\,\le}$aa} for Basic Graphs}
\label{section:aa}

\tcsautomoveaddto{main}{
  \subsection{Proofs for Section~\ref{section:aa}}
}

The classification of the complexity of $E_1^\le p$ also changes when we
restrict the admissible input structures to be basic graphs:
$\PClass{FD}_{\mathrm{basic}}(E_1^{\le}aa) \subseteq \Para\Class{AC}^{0}$ holds
by Lemma~\ref{lemma:basic-aa-in-ac0} and, once more, this is the only change.
Proving the lemma will be considerably easier than in the previous section, but
still demanding. Before we start, we summarize the resulting landscape for
completeness. Note that all bounds other than the just-mentioned new upper bound
have already been shown in Lemma~\ref{lemma:arb-min}.

\begin{theorem}[Trichotomy for $\PClass{FD}_{\mathrm{basic}}(E_1^{\le} p)$]
  Let $p$ be a pattern.
  \begin{enumerate}
    \item $\PClass{FD}_{\mathrm{basic}}(E_1^{\le}p) \subseteq \Para\Class{AC}^{0}$ if $p \preceq e^*a$ or $p \preceq aa$.
    \item $\PClass{FD}_{\mathrm{basic}}(E_1^{\le}p) \subseteq \Para\Class{AC}^{0\uparrow}$ but $\PClass{FD}_{\mathrm{basic}}(E_1^{\le}p) \not\subseteq\Para\Class{AC}^{0}$, if $aaa \preceq p$ or $eaa \preceq p$, and $p\preceq e^*a^*$.
    \item $\PClass{FD}_{\mathrm{basic}}(E_1^{\le}p)$ contains a $\Class{W[1]}$-hard problem, if $ae\preceq p$.
  \end{enumerate}
\end{theorem}

\begin{lemma~}\label{lemma:basic-aa-in-ac0}
  $\PClass{FD}_{\mathrm{basic}}(E_1^{\le}aa) \subseteq \Para\Class{AC}^0$.
\end{lemma~}

\begin{proof~}
  As in the previous section, we can reuse some ideas from the
  literature, but need to take care of some extra complications caused
  by the need to limit the sizes of the solution sets. In particular,
  we will use the notion of \emph{cardinality constraints} introduced
  in~\cite{Tantau15} for the study of the $E_1aa$ case: For two sets
  $C,D \subseteq \{0,1,2\}$ define $\PLang{csp}^{\le}\{C,D\}$ as
  follows. The instances for this 
  problem consist of a finite universe~$U$, a function~$P$
  that maps \emph{each} two-element subset $\{x,y\} \subseteq U$ to
  either $C$ or~$D$ (so, unlike normal constraint satisfaction
  problems, a constraint must be stated for every single pair of
  variables), and a number~$k$.  A \emph{solution} for $P$ is a
  subset $X \subseteq U$ of size $|X| \le k$ such that for all
  two-element subsets $\{x,y\} \subseteq U$ we have $\left|\{x,y\}
    \cap X\right| \in P(x,y)$.
  We call $\bigl(U, P^{-1}(C)\bigr)$ the \emph{$C$-graph
    of~$P$} and note that this is just the set of edges that are mapped
  to $C$ by~$P$. The \emph{$D$-graph} is defined as $\bigl(U,
    P^{-1}(D)\bigr)$; and observe that every two-element set  belongs to
  exactly one of these two graphs except when $C =D$ in which case
  both graphs are the complete cliques.
  It is shown in \cite[Lemma 3.1]{Tantau15} that all problems in
  $\Class{FD}(Eaa)$ reduce to $\Lang{csp}\{C,D\}$ (without the ``$\le
  k$'' restrictions) for some $C$ and~$D$. We need to following
  variant:
  \begin{claim}
    All problems in $\PClass{FD}(E_1^\le aa)$ reduce to
    $\PLang{csp}^{\le}\{C,D\}$ for some $C$ and $D$ via $\Para\Class{AC}^0$
    reductions.  
  \end{claim}
  \begin{proof}
    The reduction is a trivial reencoding in which the solutions of
    the \textsc{csp} instances are exactly the sets that satisfy the
    formula when assigned to the existentially bound second-order
    variable. In particular, solutions and assigned sets have the same
    sizes and satisfy the same size restrictions.
  \end{proof}
  It remains to show $\PLang{csp}^{\le}\{C,D\} \in \Para\Class{AC}^0$
  for all $C$ and~$D$. Let us go over the possible choices in a case
  distinction. In each case, we assume that we are given a
  set~$U$, a function~$P$, and a number~$k$ and must decide whether
  $P$ has a size-$k$-or-less solution.

  \subparagraph*{The case $C = \emptyset$.} As the constraint $C$
  cannot be satisfied, if it is present at all (which is trivial to
  check), there cannot be a solution. So assume that the $C$-graph is
  empty and, thus, the $D$-graph is a complete clique.
  If $0 \in D$, then $X = \emptyset$ is always a solution and we can
  simply accept. Otherwise, three cases remain: $D = \{1\}$, $D =
    \{2\}$, and $D = \{1,2\}$.

  For $D = \{1\}$, any solution of $P$ is a
  2-coloring of the
  $D$-graph -- and such a solution clearly exists (as $D$ is a clique)
  iff $|U| \le 2$.

  For $D = \{2\}$, the only allowed solution is $X =
    U$ (for $|U| \ge 2$), so the question is whether $|U| \le k$ holds
  (this can be checked in $\Para\Class{AC}^0$).

  Finally, for $D =
    \{1,2\}$, we ask whether there is a solution $X \subseteq U$ such
  that for any pair $\{x,y\} \subseteq U$ at least one of $x$ and $y$
  belongs to~$X$. This is the same as asking whether all but at most
  one of the elements of~$U$ are in~$X$. In other words, we must check
  whether $|U| - 1 \le k$ holds, which is again possible in
  $\Para\Class{AC}^0$.

  \subparagraph*{The cases $C = \{0\}$, $C=\{0,1\}$, and
    $C=\{0,1,2\}$.} The case $D = \emptyset$ needs
  not be covered as this was already covered in the previous case with
  the roles of $C$ and $D$ exchanged. Also as in the first case, when
  $0 \in D$, the solution $X = \emptyset$ is always valid. So, again,
  consider the three possible cases $D = \{1\}$, $D=\{2\}$, and $D =
    \{1,2\}$.

  In all three cases, $X$ must (at least) form a vertex cover of the
  $D$-graph in order to be a solution. This allows us to apply the
  following standard kernelization method: Look for high-degree
  vertices in the $D$-graph. Here, ``high degree'' means degree more
  than $k$ since such vertices \emph{must} be in any size-$k$ vertex
  cover of the $D$-graph. If there are more that $k$ such vertices
  (which can be checked in $\Para\Class{AC}^0$), no solution can
  exist. Otherwise, mark these vertices as members of~$X$. Next, if
  there are more than $\binom{k}{2}$ edges in the $D$-graph that do
  not contain any high-degree vertices, we also know that no solution
  of size~$k$ can exist. Otherwise, we iterate over all possible ways
  of choosing $X$ restricted to the vertices that are not only
  connected to high-degree vertices in the $D$-graph (there can only
  be $k^2$ many of them). The high-degree vertices are, of course,
  also in~$X$, while all other vertices are not in~$X$. For each
  choice, we check whether it is a solution and, if so, accept (this
  is correct as in the $C$-graph not choosing a vertex is never wrong).

  \subparagraph*{The case $C = \{0,2\}$.} We can ignore $D =
    \emptyset$, $D = \{0\}$, $D=\{0,1\}$, and $D=\{0,1,2\}$, as these
  are already covered by the previous cases with the roles of $C$ and~$D$
  exchanged. We can also ignore $D = \{0,2\}$ as this already covered by
  $C = \emptyset$ and $D = \{0,2\}$. Thus, the remaining cases are,
  again, $D = \{1\}$,  $D=\{2\}$ and $D = \{1,2\}$.

  We can once more kernelize as in the previous case. Let $H \subseteq
    U$ be the set of high-degree vertices and let $M \subseteq U$ be the
  vertices that are not only connected to high-degree vertices in the
  $D$-graph. Let $L = U \setminus H \setminus M$ be the remaining
  vertices. Then any solution $X$ must contain all of~$H$ as well as
  some subset $M' \subseteq M$. Since $|M| \le k^2$, we can iterate
  over all possible choices for~$M'$. Our objective is to find
  (``extend'') for each $M'$ an $L' \subseteq L$ such that if there is
  a solution at all, then for an appropriate $M'$ the set $X\coloneq H \cup
    M' \cup L'$ will be a solution.

  To decide whether $u \in L$ is a member of $L'$ or not, we consider
  the edges in the $C$-graph between $u$ and $v \in H \cup M$. If such
  an edge exists, we put $u$ into $L'$ if $v \in H \cup M'$ (since $C
    = \{0,2\}$ enforces that the membership in $X$ is the same for $u$
  and $v$). If there are conflicting values ($u$ is both connected to
  $v$ and $v'$, one being member of $H \cup M'$ and the other not), we
  know $M'$ cannot be extended to a solution anyway. The remaining
  case is that $u$ is connected to all $v \in H \cup M$ via edges of
  the $D$-graph. For $D = \{1\}$ and $D=\{2\}$ this also immediately
  enforces or prohibits membership of $u$ in~$L'$. The remaining case
  is $D = \{1,2\}$.

  As $u$ is not a member of $M$, we conclude that $u$ is connected
  only to vertices in~$H$. As $u$ is not a member of~$H$, we also
  conclude that $u$ has degree at most~$k$. Let $I$ be the set of all
  such vertices. As just observed, in the $D$-graph these vertices are
  all connected to vertices in~$H$, meaning that regardless of whether
  or not we put them into $L'$, the $D$-constraint will be satisfied
  as all vertices in $H$ are part of~$X$. Now consider any $u \in I$
  and any $v \in L \setminus I$. Then $P(\{u,v\}) = C$ must hold
  since, otherwise, we would have had $u \in M$. Since for the
  vertices in $L \setminus I$ we have already determined membership,
  we immediately get membership for~$u$ also. Finally, if $L \setminus
    I$ is empty, we simply put no element of $I$ into $X$, which is
  correct, as we can choose membership freely for the elements and for
  a minimization problem it does not hurt to leave out elements.

  \subparagraph*{The case $C = \{1\}$.} We can ignore both $D =
    \emptyset$ and all cases with $0 \in D$, as these are already
  covered by the previous cases with the roles of $C$ and~$D$
  exchanged. We can also ignore $D = \{1\}$ as this already covered by
  $C = \emptyset$ and $D = \{1\}$. Thus, the remaining cases are
  $D=\{2\}$ and $D = \{1,2\}$.

  The argument is now mostly the same as in the previous case
  $C=\{0,2\}$. The only difference is at the end when $L'$ is
  determined: When $P(\{u,v\}) = C$ holds for some $u \in I$ and $v
    \in L \setminus I$, the induced membership in $L'$ is now the exact
  opposite as before. The rest of the argument is the same.

  \subparagraph*{The case $C = \{2\}$.} As before, most choices for
  $D$ are already dealt with. The only remaining ones are $D = \{1\}$
  and $D = \{1,2\}$. However, we can argue as in the previous two
  cases, only now the constraint $C = \{2\}$ actually enforces $L' =
    L$, except when $L = I$ and $|I| = 1$ and, then $L = \emptyset$ can
  be used.

  \subparagraph*{The case $C = \{1,2\}$.} This has already been taken
  of for all choices of $D$.
\end{proof~}

\section{Conclusion}
\label{section:conclusion}

We gave a complete characterization of the tractability
frontier of weighted \textsc{eso} logic over basic graphs, undirected graphs, and
arbitrary structures. While in some cases our results mirror the
classical complexity landscape, other cases yield clearly different
results. The proofs differ significantly from the classical
setting and make extensive use of tools from parameterized complexity
theory. Especially for the class 
$\PClass{FD}_{\mathrm{basic}}(E^{\ge} ae)$, sophisticated machinery
is needed to establish the upper bound. 
Whether we require solutions to have size exactly~$k$, at most~$k$, or at
least~$k$ plays a central role in the complexity of the describable problems.
While the class $\PClass{FD}_{\mathrm{basic}}(E^{\ge} ae)$ can be shown to be
included in $\Para\Class{AC}^0$, the classes $\PClass{FD}_{\mathrm{basic}}(E^{=}
ae)$ and $\PClass{FD}_{\mathrm{basic}}(E^{\le} ae)$ both contain
$\Class{W}[2]$-hard problems. Similarly, while
$\PClass{FD}_{\mathrm{basic}}(E^{\le} aa)$ is contained in $\Para\Class{AC}^0$,
both $\PClass{FD}_{\mathrm{basic}}(E^{=} aa)$ and
$\PClass{FD}_{\mathrm{basic}}(E^{\ge} aa)$ contain $\Class{W}[1]$-hard problems. 

An obvious further line of research is to consider the prefixes $E_i^= p$,
$E_i^\ge p$, and $E_i^\le p$ for $i \ge 2$, that is, the non-monadic case, and
also multiple monadic quantifiers. While in the classical setting it turns
out~\cite{GottlobKS04} that we can normally reduce non-monadic quantifiers to
(possibly multiple) monadic ones, it is not clear whether the same happens in
the parameterized setting. Just pinpointing the complexity of, say,
$\PClass{FD}_{\mathrm{basic}}(E_2^\ge ae)$ seems difficult. 

A different  line of inquiry is to further investigate the
patterns that lead to intractable problems: In the unweighted setting,
all \textsc{eso}-definable problems lie in $\Class{NP}$ and 
if the class is not contained in~$\Class P$, then it contains an
$\Class{NP}$-complete problem. Our   
intractability results range from $\Class{W}[1]$-completeness to
$\Para\Class{NP}$-completeness. Can we find for every~$t$ a pattern for
which we get classes that contain $\Class{W}[t]$-hard problems and are
contained in~$\Class{W}[t]$? 

Our results also shed some light on \emph{graph modification
problems,} where we have a fixed first-order formula 
$\phi$ and are given a pair $(G,k)$. The objective is to modify the
graph as little as possible (for instance, by deleting as few vertices
as possible) such that for the resulting graph $G'$ we have $G'\models
\phi$. Fomin et al.~\cite{FominGT20} have recently
shown a complexity dichotomy regarding the number quantifier
alternations in~$\phi$. Since it is not difficult to encode the
``to be deleted vertices'' using a $\exists^\le D$ quantifier, at
least the upper bounds from our paper also apply to vertex deletion
problems. We believe that our results can be extended to also cover
lower bounds and, thereby, to give exact and complete classifications
of the complexity of vertex deletion problems in terms of the
quantifier pattern of~$\phi$.

\bibliography{main}

\tcsautomoveinsert{main}

\end{document}